\documentclass[11pt,twoside,a4paper]{article}
\usepackage{pgfplots}\pgfplotsset{/pgf/number format/use comma,compat=newest}
\usepackage{amsmath,amssymb,amstext}
\usepackage{amsthm,bm}
\usepackage{epsfig,endnotes}
\usepackage{setspace}
\usepackage{xcolor}
\usepackage{graphics}
\usepackage{graphicx}
\usepackage{multirow}
\usepackage{booktabs}
\usepackage{geometry}
\usepackage{fancyhdr}
\usepackage{hyperref}
\usepackage{boxhandler}
\usepackage[font=small,labelfont=bf]{caption}
\usepackage{subcaption}
\usepackage{epsfig,endnotes}
\usepackage[romanian,english]{babel}
\usepackage{lscape}
\usepackage{combelow}
\usepackage{authblk}
\usepackage{geometry}
\usepackage{cite}
\usepackage[normalem]{ulem}
\usepackage[font=footnotesize,labelfont=bf]{caption}
\newenvironment{sistema}{\left\lbrace\begin{array}{@{}l@{}}}{\end{array}\right.}

\providecommand{\keywords}[1]
{
	\small	
	\textbf{\textit{Keywords---}} #1
}

\title{Displaying risk in mergers: a diagrammatic approach for exchange ratio determination}
\author[1]{Alessandra Mainini}
\author[2]{Enrico Moretto$^{\ast,}$}
\author[2]{Daniela Visetti}

\affil[1]{Dipartimento di scienze economiche e sociali, Universit\`a Cattolica del Sacro Cuore, Piacenza, Italia}
\affil[2]{Dipartimento di economia, metodi quantitativi e strategie di impresa, Università di Milano-Bicocca, Milano, Italia\\ 
	
	$^{\ast}$corresponding author: \href{mailto:enrico.moretto@unimib.it}{enrico.moretto@unimib.it}}

\date{}
\begin{document}
	\maketitle
	\abstract{This article extends, in a stochastic setting, previous results in the determination of feasible exchange ratios for merging companies. A first outcome is that shareholders of the companies involved in the merging process face both an upper and a lower bounds for acceptable exchange ratios.
		Secondly, in order for the improved `bargaining region' to be intelligibly displayed, the diagrammatic approach developed by Kulpa is exploited.
		
		\keywords{Mergers and acquisitions, exchange ratio determination, synergy, risk-adjusted performance, diagrammatic representation}
		
		
		\section{Introduction, literature review and motivation}\label{Intro}
		
		Mergers and acquisitions have been, and still are a widely studied topic in financial literature under both a theoretical and an empirical point of view. A concise summary of contributions in this field is virtually impossible; recent attempts are King et al. \cite{king2018mergers} and Risberg et al. \cite{risberg2015routledge}.
		
		Companies merge for various reasons, but with a unique goal: the creation of synergy, that is a, hopefully positive, change in the  behavior of the equity of the resulting company (here denoted by $M$) when compared to those of the equities of the preexisting ones, namely the acquiring, $A$, and the acquired, or target, $B$. 
		
		Relevant contributions to the financial evaluation of synergies  are due to Gupta and Gerchak \cite{gupta2002quantifying}, Leland \cite{leland2007financial}, and are broadly summarized in Chapter 15 of Damodaran \cite{damodaran2008damodaran}. 
		
		This paper focuses on the impact of synergy for shareholders of the merging companies and extends, in a stochastic environment, known results (Larson and Gonedes, \cite{larson1969business} and Yagil \cite{yagil1987exchange}; in what follows, this approach is referred to as LG-Y).
		
		The attention of this article resides on {\it stock-for-stock agreements}, where shareholders of the target company swap, if the merger is finalized, $r$ stocks of company $M$ for each share of company $B$ they own. Shareholders of the acquiring company $A$ receive, instead, one stock of $M$ for each share they own. Stockholders of either company $A$ or $B$ become shareholders of company $M$.
		
		A topic that has been almost totally neglected in financial research is the fair determination of the exchange ratio (ER) $r$.
		In a deterministic framework, the LG-Y setting is the only known contributions. Here, companies are evaluated using the  the dividend discount model (Gordon and Shapiro, \cite{gordon1956capital}). The result is the explicit determination of the bargaining region (BR), that is the range for exchange ratios that do not reduce shareholders' wealth. 
		
		A generalization with a dynamical extension of the LG-Y setting  is due to Cigola and Modesti \cite{cigola2008note} while a recent contribution, that encompasses a number of accounting observations, is in Taliento \cite{taliento2023valuation}.
		
		As noted by Amihud and Lev \cite{amihud1981risk}, risk reduction can be a driver pushing companies to merge. An attempt to introduce risk in the determination of the exchange ratio has been proposed by Moretto and Rossi \cite{moretto2008exchange}. This contribution, though, does not identify a BR; it provides, instead, a unique risk-corrected ER. 
		
		More recently, Toll and Hering \cite{toll2017valuation} analyze, in a stochastic context, the effects of a merger for shareholders by means of expected utility theory.
		
		Entering the realm of randomness in financial markets requires a solid theory for risk measures. The fundamental contribution in this context is provided by Artzner et al. \cite{artzner1999coherent}; here riskiness is measured by means of functions 
		that translate random variables into a cash amount that, if positive, has to be paired to the risk in order to make it acceptable. 
		
		In this contribution, no specific risk measure is chosen. Our analysis starts by mimicking results in LG-Y replacing the deterministic result with expected values of random stock prices and measures changes in risk the merger grants to stockholders.
		
		In the LG-Y setting, in terms of ER determination  shareholders of the acquiring and acquired companies have clear-cut and opposite interests: the larger $r$, the larger the stake of company $M$ $B$'s shareholders own, and the larger the portion of synergy they receive. This is the reason why stockholders of company $A$ aim at keeping $r$ as small as possible. Our contribution shows that, when a risk measure is introduced, something completely different and less straightforward occurs. 
		
		As a matter of fact, imposing bounds on expected values and risk measure leads to more accurate and realistic BRs.
		
		In finance, nothing comes for free. Adding a second dimension  embeds into the stochastic framework a drawback: BRs cannot be easily depicted, unless some appropriate methodology is applied. Luckily, the `diagrammatic' representation of intervals developed by Kulpa (see \cite{kulpa1997diagrammatic}, \cite{kulpa2001diagrammatic}, and \cite{kulpa2006diagrammatic}) permits to transform intervals into points on a two-dimension Cartesian plane and solve the issue. 
		
		The paper is organized as follows: Section \ref{MeAr} describes the theoretical framework on which the extension in a stochastic fashion of the LG-Y model abides. Section \ref{Kulpa} introduces Kulpa's diagrammatic representation, translates BRs in legible and easy-to-understand plots, and provides some financial insights. Section \ref{Conclusions} concludes
		
		\section{Arithmetic for merger agreements} \label{MeAr}
		\subsection{The deterministic benchmark: Larson and Gonedes, and Yagil setting} \label{SubMeAr}
		
		For the reader's sake, in this subsection we summarize the main findings of contributions from Larson and Gonedes, and Yagil (LG-Y from now on). Their approach is deterministic 
		and highlights conditions under which shareholders of the acquiring ($A$) and acquired ($B$) companies are better off once the merger agreement is settled, leading to the birth to the resulting company $M$.
		The LG-Y setting assumes that, in a {\it stock-for-stock} agreement, each stock of company $A$ translates into one stock of company $M$ while each stock of company $B$ 
		becomes $r$ stocks of company $M$, where $r \ge 0$ is the exchange ratio (ER). 
		
		{\it Stock-for-stock} refers to the exchange of stocks of the resulting with stocks of the acquired company at a predetermined conversion rate. Companies that pay for their acquisitions with stock share both the value and the risks of the transaction with the shareholders of the company they acquire. Nevertheless, as mentioned above, the LG-Y setting
		is purely deterministic, and stockholders share only the value of the transaction.
		
		Let $p_i$, $N_i$, and $\mu_i = p_i \cdot N_i$ be, respectively, the stock price, the number of outstanding stocks, and the equity value for companies $i = A, B, M$. The number of $M$'s outstanding stocks is
		$$N_M = N_A + r N_B.$$ 
		
		A merger creates some positive synergy when the equity value of $M$ corresponds at least to the sum of equity values of $A$ and $B$:
		$$\mu_M \ge \mu_A + \mu_B.$$ 
		
		A relevant quantity 
		is ratio
		\begin{equation}
			r_{\ast} = p_B/p_A\text{.} \label{rast}
		\end{equation}
		which represents the relative value of a stock of company $B$ when compared with the value of one stock of company $A$.
		
		The LG-Y approach provides an explicit expression for the bounds of the {\it bargaining region} (BR), that is the interval containing all ERs for which shareholders of companies $A$ and $B$ simultaneously experience an increase in their equity values and, thus, might accept to conclude the merger. 
		
		Existence and magnitude of this interval depend on synergy created by the merger.
		
		Company $A$'s stockholders enjoy a benefit from this agreement when the price of stocks they own after the merger is equal, at least, to the price of stocks they previously owned, 
		that is
		\begin{equation*}
			p_M = \frac{\mu_M}{N_A + r N_B} \ge \frac{\mu_A}{N_A} = p_A\text{.} 
		\end{equation*}
		Simple algebra yields an upper bound for $r$:
		\begin{equation}
			r \le \frac{\mu_M - \mu_A}{\mu_A} \cdot \frac{N_A}{N_B} = r_{\ast} \cdot \frac{\mu_M - \mu_A}{\mu_B} = \overline{r}_{\mu}\left(\mu_M\right). \label{upev}
		\end{equation}
		Condition (\ref{upev}) denotes the largest ER company $A$'s shareholders accept.
		
		
		Company's $B$ shareholders, instead, obtain an increment in their wealth when
		\begin{equation*}
			r \cdot p_M = 
			r \cdot \frac{\mu_M}{N_A + r N_B} \ge \frac{\mu_B}{N_B} = p_B\text{.} 
		\end{equation*}
		This condition is equivalent to a lower bound for $r$:
		\begin{equation}
			r \ge r_{\ast} \cdot \frac{\mu_A}{\mu_M - \mu_B} = \underline{r}_{\mu}\left(\mu_M\right)\text{.} \label{lpev}
		\end{equation}
		
		Bounds \eqref{upev} and \eqref{lpev} depend 
		on the equity value of $M$. Further,  
		it is immediate to see that, if $\mu_M\geq\mu_A+\mu_B$, then $\underline r_\mu(\mu_M)\leq\overline{r}_\mu(\mu_M)$. In this case, the above bounds identify set
		\begin{equation}
			\mathcal{BR}_{\mu}\left(\mu_M\right) = [\underline r_{\mu}(\mu_M),\overline r_{\mu}(\mu_M)]. \label{interval_mu}
		\end{equation}
		
		The absence of synergy (i.e. $\mu_M = \mu_A + \mu_B$) shrinks 
			set $\mathcal{BR}_{\mu}\left(\mu_M\right)$ } into a singleton: $\mathcal{BR}_{\mu}\left(\mu_A + \mu_B\right) =\left\{r_{\ast}\right\}$. This expression highlights the importance of ratio \eqref{rast}.
		
\begin{figure}
			\centering
			\includegraphics[width=.55\linewidth]{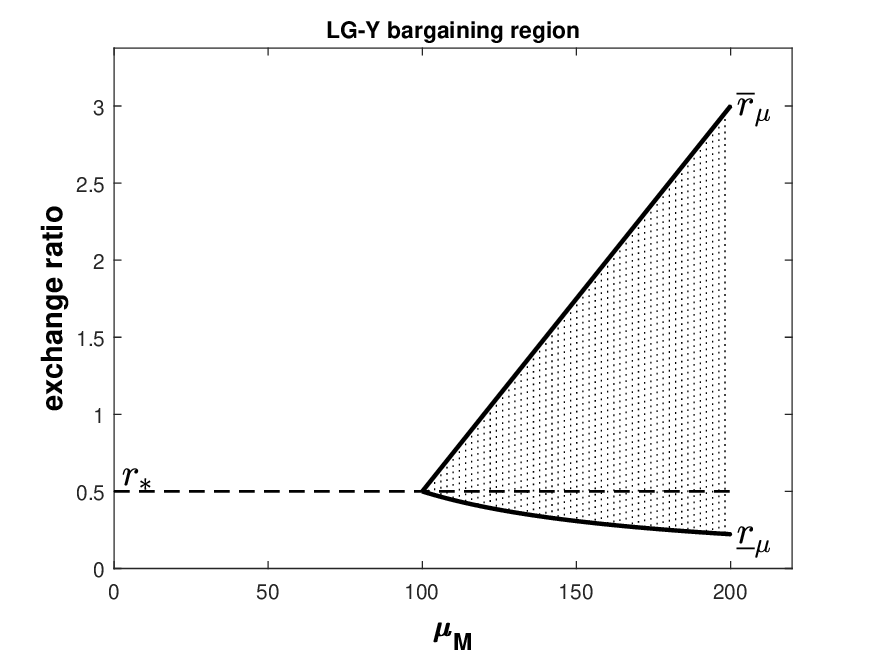}
			\caption{Graphical representation of a BR (shaded area) in the LG-Y's setting - $p_A = 4$, $p_B = 2$, $N_A = 20$, $N_B = 10$, $r_{\ast}=0.5$,  $\mu_A + \mu_B = 100$, $\mu_M \in \left[100,200\right]$.} \label{Figura-1}
		\end{figure}

		Shaded region in Figure \ref{Figura-1} represents set \eqref{interval_mu} in terms of $\mu_M$.
		
		
		
		\subsection{A stochastic extension} \label{SubMeAr-2}
		
		As mentioned above, in a stock-for-stock merger companies not only share the change in expected equity values but, also, the change in risk created by this 
		transaction.
		
	
	
	In order to expand the LG-Y results in a stochastic environment, we need to introduce and measure risk. 
	
	Randomness is represented by (non-negative) random variables $\tilde p_i$, $i = A, B, M$, that describe stock prices. We make no {\it a priori} choice on 
	their distribution, as long as expected values $m_i$ assumed to be a measure of performance and homogeneous measures of risk $\varphi_i$, in the sense of 
	Artzner et al. (see \cite{artzner1999coherent}), for $\tilde p_i$, exist. We also assume that $\varphi_i \ge 0$ so that stocks are considered a non-
	acceptable risk and shareholders might be interested in reducing it.
	
	
	
	For the sake of notational simplicity, we decide to set $m_i = p_i$ so that formulas (\ref{upev}) and (\ref{lpev}) are recovered. 
	This claim is easy to justify; the deterministic approach corresponds to the one presented in this subsection if expected values only are considered.
	
	Similarly to expression (\ref{rast}), we stress the importance of ratio
	\begin{equation}
		r_{\ast\ast} = \varphi_B/\varphi_A, \label{rastast}
	\end{equation}
	that expresses the relative risk 
		of a share 
	of company $B$ w.r.t. riskiness of a stock of company $A$.
	
	Introducing a risk measure {for equities allows to define ratio $ \lambda_i = p_i/\varphi_i$, 
		that is the coefficient of variation of random variable $\tilde{p}_i$ as well as a relative value that expresses a stock's risk-corrected measure of 
		performance.  
		
		Condition $\lambda_A > \lambda_B$ (respectively $\lambda_A < \lambda_B$), that is the risk-corrected performance of stock $A$ is better (respectively worse) 
		than $B$'s, is equivalent to $r_{\ast\ast} > r_{\ast}$ (respectively $r_{\ast\ast} < r_{\ast}$).
		

		
		
		Let $\rho_i = \varphi_i \cdot N_i$. Shareholders of company $A$ benefit of a smaller risk after the merger 
		when 
		\begin{equation}
			\varphi_M = \frac{\rho_M}{N_A + rN_B}\leq \frac{\rho_A}{N_A} = \varphi_A.
		\end{equation}
		This inequality yields a lower bound for ERs:
		
		\begin{equation}
			r \ge r_{\ast\ast}\cdot\frac{\rho_M - \rho_A}{\rho_B} = \underline r_{\rho}(\rho_M). \label{Cond1SD}
		\end{equation}
		Here, to prevent $r$ from becoming negative, condition $\rho_M > \rho_A$ must hold.
		
		Stockholders of company $B$ obtain a reduction in their equity risk when 
		\begin{equation}
			r\cdot \varphi_M = r\cdot\frac{\rho_M}{N_A + rN_B}\leq\frac{\rho_B}{N_B}=\varphi_B\text{.}
			\label{Cond2SDbis}
		\end{equation}
		This inequality leads to an upper bound for exchange ratios:
		\begin{equation}
			r \le r_{\ast\ast}\cdot\frac{\rho_A}{\rho_M - \rho_B} = \overline{r}_{\rho}(\rho_M)
			\label{Cond2SD}
		\end{equation}
		where imposing $\rho_M > \rho_B$ forbids $r$ to become negative. 
		
		Bounds (\ref{Cond1SD}) and (\ref{Cond2SD}) yield positive ERs when $\rho_M > \max\left\{\rho_A,\,\rho_B\right\}$. This constraint implies that the reduction 
		in riskiness a merger can achieve reduces the risk of company $M$ at most up to the largest risk of the merging ones.
		
		
		
		
		Inequalities (\ref{Cond1SD}) and (\ref{Cond2SD}) identify a second bargaining region that is non empty whenever $\rho_M \le \rho_A + \rho_B$ and contains positive ERs if $\rho_M > \max\left\{\rho_A,\rho_B\right\}$. These conditions lead to 
		\begin{equation}
			\max\left\{\rho_A,\rho_B\right\} < \rho_M \le \rho_A + \rho_B
			\label{range-rho}
		\end{equation}
		while this second set is
		\begin{equation}
			\mathcal{BR}_{\rho}\left(\rho_M\right) = \left[\underline{r}_{\rho}(\rho_M),\overline{r}_{\rho}(\rho_M)\right] \label{interval_rho}
		\end{equation}
		for $\max\left\{\rho_A,\,\rho_B\right\} <  \rho_M \leq \rho_A + \rho_B$.
		Interval (\ref{interval_rho}) collapses into a singleton when the merger does not reduce 
		riskiness, that is $\rho_M = \rho_A + \rho_B$. In fact, $ \mathcal{BR}_{\rho}\left(\rho_A + \rho_B\right) = \left\{r_{\ast\ast}\right\}$.
		
		
		A comparison between inequalities (\ref{upev}) and (\ref{Cond1SD}), that relate to company $A$'s stockholders, and between inequalities (\ref{lpev}) and 
		(\ref{Cond2SD}), that identify which exchange ratios are acceptable for company $B$'s stockholders, leads to a counterintuitive as well as interesting 
		outcome. 
		
		If bound (\ref{upev}) suggests that stockholders of company $A$ want the exchange ratio to be as small as possible as this would grant them a large stake of expected synergy, bound (\ref{Cond1SD}) 
			indicates 
		that, if the number of $M$'s outstanding stocks is not large enough, the risk each stock ends up carrying is not less than the risk of each $A$'s stock.
		
		A similar line of reasoning can be done for company $B$'s stockholders. Bound (\ref{lpev}) implies that shareholders of the acquired company want to fix the exchange ratio as large as possible because, in this case, they seize a large portion of expected synergy. Bound (\ref{Cond2SD}), though, reveals that acceptable ERs cannot exceed a given threshold. If this happens, the stocks these stockholders obtain in exchange of each share they owned carry larger risk\footnote{The trivial arithmetic explanation of this claim is in Appendix \ref{App1}.}.

		The existence of exchange ratios upper and lower bounds for  both categories of shareholders permits to introduce two more sets that we call {\it consistency region} (CR). These intervals, when non-empty, contain all ERs that simultaneously grant both an increase in expected equity values and a reduction in equity risk for both groups of shareholders.
		
		These sets are defined as
		\begin{equation}
			\mathcal{CR}_{\text{A}}\left(\mu_M,\rho_M\right) = \left[\underline{r}_{\rho}\left(\rho_M\right), \overline{r}_{\mu}\left(\mu_M\right)\right] \label{CR-A}
		\end{equation}
		with $\mu_M \ge \mu_A + \mu_B$ and $\max\left\{\rho_A,\,\rho_B\right\} <  \rho_M \leq \rho_A + \rho_B$ for shareholders of company $A$, and 
		\begin{equation}
			\mathcal{CR}_{\text{B}}\left(\mu_M,\rho_M\right) = \left[\underline{r}_{\mu}\left(\mu_M\right),\overline{r}_{\rho}\left(\rho_M\right)\right] \label{CR-B}
		\end{equation}
		with $\mu_M \ge \mu_A + \mu_B$ and $\max\left\{\rho_A,\,\rho_B\right\} <  \rho_M \leq \rho_A + \rho_B$
		for those of company $B$.
		
		Assuming a merger yields an improvement in terms of both larger expected wealth and smaller risk might seem a rather `too-good-to-be-true' requirement. Still, the key financial notion of diversification is based on better performance paired with less risk. 
		
		Interval (\ref{CR-A}) is non-empty
		when
		$$
		\frac{\mu_M - \mu_A}{\rho_M - \rho_A} \ge \frac{p_A} {\varphi_A} = \lambda_A\text{.} 
		$$
		To proceed with this analysis, let $s \ge 0$ be the expected synergy and $v \ge 0$ the reduction in risk, so that $\mu_M = \mu_A + \mu_B +s$ and $\rho_M = \rho_A + \rho_B 
		-v$}. The above condition becomes
	$$
	\frac{\mu_B + s}{\rho_B - v} \ge \lambda_A\text{.} 
	$$
	
	
	Consider, at first, the limit case where $s=0$ and $v=0$ (that is, the merger carries no benefit whatsoever). Interval (\ref{CR-A}) is non-empty if
	$$
	\frac{\mu_B}{\rho_B} = \lambda_B \ge \lambda_A\text{,} 
	$$
	that is when company $B$ is preferable to $A$ in terms of risk-corrected performance. Shareholders of company $A$ are favourable to complete the merger regardless of its positive 
	synergy creation and/or risk reduction because company $M$ has an improved risk-corrected performance when compared $A$'s one.  
	
	The trivial algebra that explains the above claim, and the one regarding company $B$ below, is relegated in Appendix \ref{App-3}.
	
	
	This, of course, holds also when $s > 0$ 
		and/or $0<v<\rho_B$ 
	as, if $\lambda_B \ge \lambda_A$,
	$$
	\frac{\mu_B + s}{\rho_B - v} > \lambda_B \ge \lambda_A\text{.}
	$$
	Further, if $\lambda_B \ge \lambda_A$ (that is $r_{\ast} \ge r_{\ast\ast}$)
	$$
	\mathcal{CR}_{\text{A}}\left(\mu_A + \mu_B,\rho_A+\rho_B\right) = \left[r_{\ast\ast},r_{\ast}\right]\text{.}
	$$
	
	Consider case $\lambda_B < \lambda_A$. Here, the no improvement due to synergy case yields 
	$$
	\mathcal{CR}_{\text{A}}\left(\mu_A + \mu_B,\rho_A+\rho_B\right) = \emptyset.
	$$
	
	The resulting company the merger creates has a risk-corrected performance which is smaller than $\lambda_A$. 
		Company A's shareholders agree on the merger only if they receive some compensation that offsets their reduction in risk-corrected performance. This occurs if
		$$
		s \ge \lambda_A\left(\rho_B -v \right) - \mu_B = \overline{s}_A\text{,}
		$$
		that is the condition for non-emptiness of set $\mathcal{CR}_A$. Here, for each  $v \in \left[0,
		\rho_B\right)$ there exists a threshold $\overline{s}_A$ that linearly decreases 
			w.r.t. 
		$v$. The above inequality 
			indicates 
		that the larger the reduction in risk, the smaller the minimum expected synergy that provides company $A$'s stockholders with a range of acceptable ERs.
		
		A similar analysis can be carried out for interval (\ref{CR-B}). This set is non-empty when 
		$$
		\frac{\mu_M - \mu_B}{\rho_M - \rho_B} = \frac{\mu_A + s}{\rho_A - v} \ge \frac{\mu_B} {\rho_B} = \lambda_B\text{.} 
		$$
		Limit case $s=0$, $v = 0$ yields $\lambda_A \ge \lambda_B$ so that company $B$'s shareholders agree on the merger even if there is no synergistic effect if company $A$ has a better risk-adjusted performance. 
		
		In this case
		$$
		\mathcal{CR}_{\text{B}}\left(\mu_A + \mu_B,\rho_A+\rho_B\right) = \left[r_{\ast},r_{\ast\ast}\right]\text{.}
		$$
		
		When, instead, $\lambda_A < \lambda_B$, set $\mathcal{CR}_{\text{B}}$ is non-empty when expected synergy  fulfills condition
		$$
		s \ge \lambda_B\left(\rho_A - v\right) -\mu_A = \overline{s}_B\text{.}
		$$

	The bargaining region in the stochastic setting is the bivariate set
	$$
	\mathcal{BR}_{\mu,\rho}\left(\mu_M,\rho_M\right) = \mathcal{BR}_{\mu}\left(\mu_M\right) \cap \mathcal{BR}_{\rho}\left(\rho_M\right)\text{.}
	$$

Set $\mathcal{BR}_{\mu,\rho}$ can be in one of six different cases, summarized in Table \ref{Table01}; four of them are non-empty. Each case corresponds to one of the sets $\mathcal{BR}$ and $\mathcal{CR}$ defined above. A detailed description of such cases is in Section \ref{Kulpa}. 

	\begin{table}
		\centering
		\begin{tabular}{r||c|c|c}
			& $\overline r_{\rho} < \underline r_{\mu} < \overline r_{\mu}$ & $ \underline r_{\mu} < \overline r_{\rho} < \overline r_{\mu}$ & $ \underline r_{\mu} < \overline r_{\mu} < \overline r_{\rho}$ \\ \hline \hline
			\multirow{2}{*}{$\underline r_{\rho} < \underline r_{\mu} < \overline r_{\mu}$} & \multirow{2}{*}{$\emptyset$} & case 1 - $\left[\underline r_{\mu},\overline r_{\rho}\right]$ & case 2 - $\left[\underline r_{\mu},\overline r_{\mu}\right]$ \\
			& & $\mathcal{CR}_B(\mu_M,\rho_M)$ & $\mathcal{BR}_{\mu}(\mu_M)$  \\ \hline
			\multirow{2}{*}{$\underline r_{\mu} < \underline r_{\rho} < \overline r_{\mu}$} & \multirow{2}{*}{NA} & case 3 - $\left[\underline r_{\rho},\overline r_{\rho}\right]$ & case 4 - $\left[\underline r_{\rho},\overline r_{\mu}\right]$ \\
			& & $\mathcal{BR}_{\rho}(\rho_M)$ & $\mathcal{CR}_A(\mu_M,\rho_M)$ \\ \hline
			$\underline r_{\mu} < \overline r_{\mu} < \underline r_{\rho}$ & NA & NA & $\emptyset$
		\end{tabular}
		\caption{Possible cases for $\mathcal{BR}_{\mu,\rho}\left(\mu_M,\rho_M\right)$. Rows and columns in this table display all possible juxtapositions of endpoints $\underline{r}_{\rho}\left(\rho_M\right)$ and $\overline{r}_{\rho}\left(\rho_M\right)$ 
				w.r.t. 
			interval $\left[\underline{r}_{\mu}\left(\mu_M\right),\overline{r}_{\mu}\left(\mu_M\right)\right]$. For sake of compactness, in the topmost row and leftmost column, variables $\mu_M$ and $\rho_M$ have been dropped. NA denotes non-attainable cases.} 
	\label{Table01}
\end{table}

Along the lines of \cite{yagil1987exchange} and \cite{larson1969business}, the issue is now the geometric representation of the bargaining region corresponding to the stochastic setting,  $\mathcal{BR}_{\mu,\rho}\left(\mu_M,\rho_M\right)$, by varying $\mu_M$ and $\rho_M$.
	Its representation is a subset of the $\mathbb R^3$ space. However, to display this set in a more intelligible way, we decide to eliminate one dimension, and represent the bargaining region in the $\mathbb R^2$ plane.

Fortunately, Kulpa does the trick, as he provides a methodology for representing intervals as points 
on the plane. His results are summarized 
and applied in the next section.  

\section{Diagrammatic representation of the bargaining region} \label{Kulpa}

In his diagrammatic approach, Kulpa (see, for example, \cite{kulpa1997diagrammatic}, \cite{kulpa2001diagrammatic}, and \cite{kulpa2006diagrammatic}) develops a useful methodology for representing bounded intervals $\left[a,b\right] \subset \mathbb{R}, a \leq b$ as points on a plane. This is done, as Kulpa writes, citing \cite{simon1996sciences}, in \cite{kulpa1997diagrammatic}, with the idea that ``\textit{solving a problem simply means representing it so as to make the solution transparent}''. 


Based on this approach, each non-empty interval of the form $\left[a,b\right]$ is set in a one-to-one correspondence with point
\[
(x,y) = \left(\frac{a + b}{2},\;\frac{b - a}{2} \right) \ \in \mathbb{R}^2
\]
where the two coordinates represent, respectively, the mid-point and the radius, or half-length, of $[a,b]$. An interval which collapses to the singleton $\{a\}$ corresponds, according to Kulpa's representation, to the point $(a,0)$, and lies on the horizontal axis.

Consider now intervals
	\begin{equation}
		\left[\underline r_{\mu}(\mu_M),\,\overline r_{\mu}(\mu_M)\right]
		\;\;\;\;\;\;\text{and}\;\;\;\;\;\;
		\left[\underline r_{\rho}(\rho_M),\,\overline r_{\rho}(\rho_M)\right],\label{intervalli_da_intersecare}
	\end{equation}
which, as described in Section \ref{MeAr}, are the sets of exchange ratios accepted by both kinds of shareholders that guarantee, respectively, an increase of the value and a decrease of the overall risk of the transaction. 

According to the diagrammatic approach, they are represented, respectively, by the 
points 
\begin{equation}
	\begin{split}
		P_{\mu}\left(\mu_M\right) & \equiv 
		\left(\frac{\underline r_{\mu}(\mu_M) + \overline r_{\mu}(\mu_M)}{2},\,\frac{\overline r_{\mu}(\mu_M) - \underline r_{\mu}(\mu_M)}{2}\right) \\
		& = \left(\frac{r_{\ast}}{2}\left(\frac{\mu_A}{\mu_M - \mu_B} + \frac{\mu_M - \mu_A}{\mu_B}\right),\;\frac{r_{\ast}}{2}\left(\frac{\mu_M - \mu_A}{\mu_B} - \frac{\mu_A}{\mu_M - \mu_B}\right)
		\right)\label{punto_iperbole_mu}
	\end{split}
\end{equation}
and
\begin{equation}
	\begin{split}
		P_{\rho}\left(\rho_M\right) & \equiv 
		\left(\frac{\underline r_{\rho}(\rho_M) + \overline r_{\rho}(\rho_M)}{2},\,\frac{\overline r_{\rho}(\rho_M) - \underline r_{\rho}(\rho_M)}{2}\right) \\
		& = 
		\left(\frac{r_{\ast\ast}}{2}\left(\frac{\rho_M - \rho_A}{\rho_B} + \frac{\rho_A}{\rho_M - \rho_B}\right),\;\frac{r_{\ast\ast}}{2}\left(\frac{\rho_A}{\rho_M - \rho_B} - \frac{\rho_M - \rho_A}{\rho_B}\right)\right)
		\label{punto_iperbole_rho}
	\end{split}
\end{equation}
	The positions of $P_{\mu}(\mu_M)$ and $P_{\rho}(\rho_M)$ in the plane vary with the admissible values of $\mu_M$ and $\rho_M$\footnote{Recall that $\mu_M$ and $\rho_M$ are admissible 
		when $\mu_M\geq \mu_A + \mu_B$ and $\max\{\rho_A,\rho_B\} < \rho_M\leq \rho_A + \rho_B$.}, thus defining two parametric curves, that we analyze in the next subsection.

	We observe that, consistently with $\mathcal{BR}_{\mu}\left(\mu_A + \mu_B\right) = \left\{r_{\ast}\right\}$ and $\mathcal{BR}_{\rho}\left(\rho_A + \rho_B\right) = \left\{r_{\ast\ast}\right\}$, 
		the corresponding points become
	$$
	P_{\mu}\left(\mu_A + \mu_B\right) = \left(r_{\ast},0\right)
	\ \ \ \ \text{ and } \ \ \ \
	P_{\rho}\left(\rho_A + \rho_B\right) = \left(r_{\ast\ast},0\right)\text{,}
	$$
	that is, the two 
	curves intersect the horizontal axis at the points $\left(r_{\ast},\,0\right)$ and $\left(r_{\ast\ast},\,0\right)$, since they represent the singletons $\{r_{\ast}\}$ and $\{r_{\ast\ast}\}$.
	
		The points $P_{\mu}\left(\mu_M\right)$ and $P_{\rho}\left(\rho_M\right)$, depending respectively on $\mu_M$ and $\rho_M$ describe the curves $\gamma(\mu_M)$ and 
		$\delta(\rho_M)$.
	
	In the next subsection (see also Appendix \ref{App2}) we show that $\gamma$ and $\delta$ represent two hyperbolas. 
	
	The explicit equations of these two curves allow us to represent the intersection of intervals in (\ref{intervalli_da_intersecare}). This achievement allows to make readable and intelligible the set of exchange ratios on which an agreement is possible.
	
	The points of each hyperbola, restricted to those satisfying the conditions imposed on $\mu_M$ and $\rho_M$, represent the set of exchange ratios accepted by all the shareholders with respect to one of the two aspects (\textit{i.e.}, expected wealth or overall risk). Since we eliminate one dimension, it becomes easier and more intuitive to determine the accepted exchange ratios by varying $\mu_M$ and $\rho_M$, as well as  those belonging to the bargaining region.
	
	
	
	\subsection{Analysis of the curves \texorpdfstring{$\gamma$}{M1} and \texorpdfstring{$\delta$}{M2}}
	
	Consider point $P_{\mu}(\mu_M)$ in \eqref{punto_iperbole_mu}. 
As proved in Appendix \ref{App2}, observing that
\begin{equation}
	x(\mu_M) - y(\mu_M) = \frac{r_{\ast}\mu_A}{\mu_M - \mu_B}, \label{x-mu-y-mu}    
\end{equation}
using trivial algebra, and removing for the sake of simplicity the dependence from $\mu_M$, the parametric curve $\gamma$ is described by hyperbola
\[
x^2 - y^2 + \alpha_1(x - y) - \alpha_2 = 0,
\]
where
\[
\alpha_1=r_{*}\left(\frac{\mu_A}{\mu_B} - 1\right)\;\;\;\text{and}\;\;\;\alpha_2=\frac{r_{*}^2\mu_A}{\mu_B}.
\]
We restrict this curve to those points that satisfy $\mu_M\geq \mu_A + \mu_B$. This condition, together with formula (\ref{x-mu-y-mu}), implies that
\[
0
< 
x(\mu_M) - y(\mu_M)\leq r_{\ast}.
\]
Therefore, the set of points in the plane that represent the exchange ratios through which stockholders of both parties share an improvement in their expected wealth is
\[
\{(x,y)\in\mathbb R^2\mid\,
x^2 - y^2 + \alpha_1(x - y) - \alpha_2 = 0\;\text{and}\;
x - r_{\ast} \leq y < x \}.
\]
The condition $y < x$ is strict because $y = x$ is an asymptote of the hyperbola 

\begin{figure}
	\centering
	\includegraphics[width=.55\linewidth]{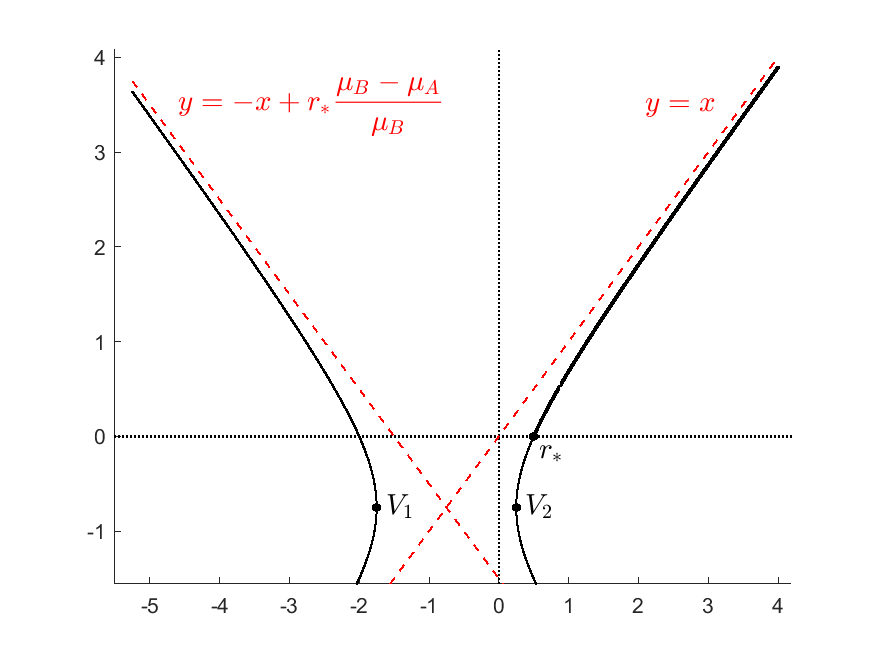}
	\caption{Kulpa's diagrammatic representation of an hyperbola depicting a $\mathcal{BR}$ - $p_A = 4$, $p_B = 2$, $N_A = 20$, $N_B = 10$, $r_{\ast}=0.5$, $V_1 \equiv\left(-1.75;-0.75\right)$, $V_2 \equiv\left(0.25;-0.75\right)$, case $\mu_A > \mu_B$. - The relevant portion of the hyperbola is the one that starts from $r_{\ast}$ and belongs to the first quadrant. The equation of the downward sloping asymptote is $y = -x -1.5$.} \label{Figura-2a}
\end{figure}

A graphical depiction of these curves is provided in Figure \ref{Figura-2a}.

Along the same lines, using the fact that
\[
x(\rho_M) - y(\rho_M) = \frac{r_{\ast\ast}(\rho_M - \rho_A)}{\rho_B},
\]
the parametric curve $\delta$ is described by
\[
x^2 - y^2 + \beta_1(x + y) - \beta_2 = 0,
\]
where
\[
\beta_1=r_{**}\left(\frac{\rho_A}{\rho_B} - 1\right)\;\;\;\text{and}\;\;\;\beta_2=\frac{r_{**}^2\rho_A}{\rho_B}.
\]
Again, we restrict the curve to the points satisfying $\max\left\{\rho_A,\,\rho_B\right\} <  \rho_M \leq \rho_A + \rho_B$. This condition yields
\[
0 < x(\rho_M) - y(\rho_M)\leq r_{\ast\ast}.
\]
Therefore, the set of points that represent the exchange ratios on which shareholders agree 
	because of the associated reduction of risk is
	\[
	\{(x,y)\in\mathbb R^2\mid\,
	x^2 - y^2 + \beta_1(x + y) - \beta_2 = 0\;\text{and}\;
	x - r_{\ast\ast} \leq y < x \}.
	\]
	
	
	\subsection{The bargaining region under Kulpa's perspective}
	
	We choose now an admissible level of post-merger risk, $\max\{\rho_A,\rho_B\} < \hat{\rho} \leq \rho_A + \rho_B$, or, in other words, we fix a point $P_{\rho}(\hat{\rho})$ along the curve $\delta$, which represents all the exchange ratios to which correspond the amount of risk $\hat{\rho}$:
	\[
	P\left( \hat{\rho}\right)
	=
	\left(\frac{\underline r_{\rho}( \hat{\rho}) + \overline r_{\rho}( \hat{\rho})}{2},\,
	\frac{\overline r_{\rho}( \hat{\rho})  - \underline r_{\rho}( \hat{\rho})}{2}\right).
	\]
	We now try to address the following questions. Under which conditions does there exist an interval of exchange ratios that makes an agreement among all the shareholders acceptable? In such a case, what is the shape of this set? Is it possible to describe its shape?
	To answer these questions, we consider the points of intersection between the interval represented by $P_{\rho}\left(\hat{\rho}\right)$, and the admissible intervals on which there is an agreement with respect to the expected wealth, that is,  those intervals represented by the points $P_{\mu}\left(\mu_M\right)$ along the curve $\gamma$, with $\mu_M\geq \mu_A + \mu_B$. Following Kulpa, the intersection between intervals $[a,b]$ and $[c,d]$ is non-empty if
	\[
	\max\{a,\,c\} \leq \min\{b,\,d\}.
	\]
	In such a case, Kulpa's representation of the intersection set is given by the point $(x,y)$, where
	\[
	x = \frac{\max\{a,\,c\}  + \min\{b,\,d\}}{2}
	\;\;\;\;\;\;\text{and}\;\;\;\;\;\;
	y = \frac{ \min\{b,\,d\} - \max\{a,\,c\}}{2}.
	\]
	Therefore, according to Kulpa's approach, the intersection between the intervals represented by  $P_{\rho}\left(\hat{\rho}\right)$ and $P_{\mu}\left(\mu_M\right)$ is non-empty if the condition
	\begin{equation}
		\max\left\{\underline r_{\mu}(\mu_M),\,\underline r_{\rho}(\hat{\rho})\right\}\leq
		\min\left\{\overline r_{\mu}(\mu_M),\,\overline r_{\rho}(\hat{\rho})\right\},\label{condizione_per_intersezione}
	\end{equation}
	is satisfied, and each point belonging to the bargaining region has $x$-coordinate 
	\[
	x(\mu_M,\hat{\rho}) = \frac{\max\left\{\underline r_{\mu}(\mu_M),\,\underline r_{\rho}(\hat{\rho})\right\} + \min\left\{\overline r_{\mu}(\mu_M),\,\overline r_{\rho}(\hat{\rho})\right\}}{2}
	\]
	and $y$-coordinate
	\[
	y(\mu_M,\hat{\rho}) =
	\frac{\min\left\{\overline r_{\mu}(\mu_M),\,\overline r_{\rho}(\hat{\rho})\right\} -  \max\left\{\underline r_{\mu}(\mu_M),\,\underline r_{\rho}(\hat{\rho})\right\}}{2}.
	\]
	We now consider all the cases that may arise from condition (\ref{condizione_per_intersezione}). For each possible scenario, we explicit what condition $\mu_M$ (respectively $\rho_M$) has to satisfy, and describe the type of intersection accordingly to Table \ref{Table01}. We refer to Appendix \ref{App-4} for the details.

\noindent \textbf{Case 1: Consistency region} $\mathcal{CR}_{B}\left(\mu_M,\hat{\rho}\right)$

\noindent Assume that $\underline r_{\mu}(\mu_M)\geq\underline r_{\rho}(\hat{\rho})$ and $\overline r_{\mu}(\mu_M)\geq \overline r_{\rho}(\hat{\rho})$. This situation is consistent, together with $\mu_M\geq \mu_A + \mu_B$, with the condition
\[
\max\left\{\dfrac{r_{\ast\ast}}{r_{\ast}}\cdot\dfrac{\mu_B\rho_A}{\hat{\rho} - \rho_B} + \mu_A,\,\dfrac{r_{\ast}}{r_{\ast\ast}}\cdot\dfrac{\mu_A(\hat{\rho} - \rho_B)}{\rho_A} + \mu_B\right\}\leq \mu_M \leq\dfrac{r_{\ast}}{r_{\ast\ast}}\cdot\dfrac{\mu_A\rho_B}{\hat{\rho} - \rho_A} + \mu_B.
\]
The same condition, written in terms of expected synergy, becomes
\[
\begin{aligned}
	\max &\left\{\mu_B\left(\dfrac{r_{\ast\ast}}{r_{\ast}}\cdot\dfrac{\rho_A}{\hat{\rho} - \rho_B} - 1\right),\,\mu_A\left(\dfrac{r_{\ast}}{r_{\ast\ast}}\cdot\dfrac{\hat{\rho} - \rho_B}
	{\rho_A} - 1\right)\right\}\leq \mu_M - \mu_A - \mu_B \\
	&\leq\mu_A\left(\dfrac{r_{\ast}}{r_{\ast\ast}}\cdot\dfrac{\mu_A\rho_B}{\hat{\rho} - \rho_A} - 1\right).
\end{aligned}
\]
The set of exchange ratios on which all shareholders agree,  represented, according to Kulpa, by the points
\[
\left( \frac{\underline r_{\mu}(\mu_M) + \overline r_{\rho}(\hat{\rho})}{2},\, \frac{ \overline r_{\rho}(\hat{\rho})- \underline r_{\mu}(\mu_M)}{2}\right),
\]
lies on the straight line of equation
\[
y = - x + \frac{r_{\ast\ast}\rho_A}{\hat{\rho} - \rho_B},
\]
which corresponds to the region  $\mathcal{CR}_{B}\left(\mu_M,\hat{\rho}\right)$ in Table  \ref{Table01}.

\begin{figure}
	\centering
	\includegraphics[width=.55\linewidth]{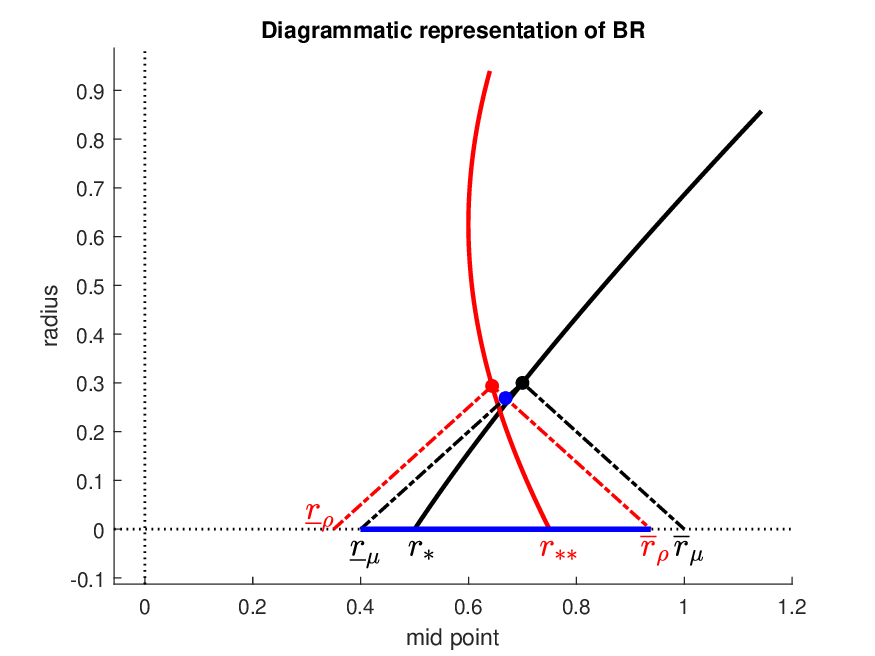}
	\caption{Kulpa's diagrammatic representation of set $\mathcal{BR}_{\mu,\rho}$ - $p_A = 4$, $p_B = 2$, $\varphi_A = 4$, $\varphi_b = 3$, $N_A = 20$, $N_B = 10$, $r_{\ast}=0.5$, $r_{\ast\ast}=0.6667$, $\mu_M = 120$, $\rho_M = 94$. The resulting interval $\mathcal{CR}_{B}$ is denoted with the blue dot and segment.} \label{Figura-5b}
\end{figure}

The diagrammatic representation for this case is in Figure \ref{Figura-5b}.

\noindent \textbf{Case 2: Bargaining  region} $\mathcal{BR}_{\mu}\left(\mu_M\right)$

\noindent Assume that $\underline r_{\mu}(\mu_M)\geq\underline r_{\rho}(\hat{\rho})$ and $\overline r_{\mu}(\mu_M)\leq \overline r_{\rho}(\hat{\rho})$. This situation is consistent, together with $\mu_M\geq \mu_A + \mu_B$, with the condition
\[
\mu_A + \mu_B \leq \mu_M \leq
\min\left\{\frac{r_{\ast}}{r_{\ast\ast}}\cdot\frac{\mu_A\rho_B}{\hat{\rho} - \rho_A} + \mu_B,\,
\frac{r_{\ast\ast}}{r_{\ast}}\cdot\frac{\mu_B\rho_A}{\hat{\rho} - \rho_B} + \mu_A
\right\}
\]
The same condition, written in terms of expected synergy, becomes
\[
0 \leq \mu_M - \mu_A - \mu_B\leq
\min\left\{\mu_A\left(\frac{r_{\ast}}{r_{\ast\ast}}\cdot\frac{\rho_B}{\hat{\rho} - \rho_A}  - 1\right),\,
\mu_B\left(\frac{r_{\ast\ast}}{r_{\ast}}\cdot\frac{\rho_A}{\hat{\rho} - \rho_B}-1\right)
\right\}.
\]
The set of exchange ratios on which all shareholders agree on lies on the curve $\gamma$, and corresponds to the region 
$\mathcal{BR}_{\mu}\left(\mu_M\right)$ in Table \ref{Table01}.

\begin{figure}
	\centering
	\includegraphics[width=.55\linewidth]{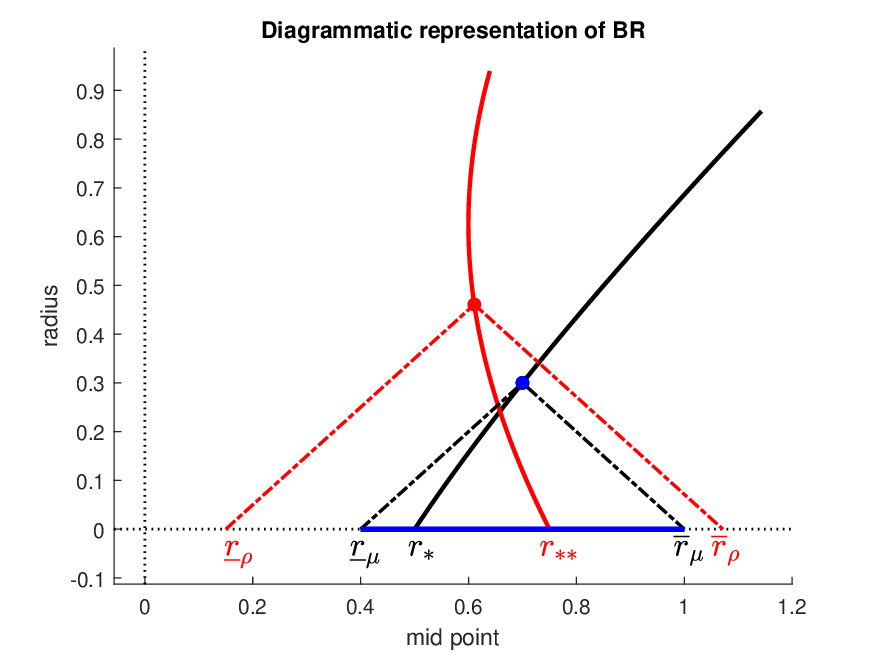}
	\caption{Kulpa's diagrammatic representation of set $\mathcal{BR}_{\mu,\rho}$ - $p_A = 4$, $p_B = 2$, $\varphi_A = 4$, $\varphi_b = 3$, $N_A = 20$, $N_B = 10$, $r_{\ast}=0.5$, $r_{\ast\ast}=0.6667$, $\mu_M = 120$, $\rho_M = 86$. The resulting interval $\mathcal{BR}_{\mu}$ is denoted with the blue dot and segment.} \label{Figura-5a}
\end{figure}

The diagrammatic representation for this case is in Figure \ref{Figura-5a}.

\noindent \textbf{Case 3: Bargaining region} $\mathcal{BR}_{\rho}\left(\hat{\rho}\right)$

\noindent Assume that $\underline r_{\mu}(\mu_M)\leq\underline r_{\rho}(\hat{\rho})$ and $\overline r_{\mu}(\mu_M)\geq \overline r_{\rho}(\hat{\rho})$. This situation is consistent, together with $\mu_M\geq \mu_A + \mu_B$, with the condition
\[
\mu_M \geq
\max\left\{\frac{r_{\ast}}{r_{\ast\ast}}\cdot\frac{\mu_A\rho_B}{\hat{\rho} - \rho_A} + \mu_B,\,
\frac{r_{\ast\ast}}{r_{\ast}}\cdot\frac{\mu_B\rho_A}{\hat{\rho} - \rho_B} + \mu_A
\right\}
\]
The same condition, written in terms of expected synergy, becomes
\[
\mu_M - \mu_A - \mu_B\geq
\max\left\{\mu_A\left(\frac{r_{\ast}}{r_{\ast\ast}}\cdot\frac{\rho_B}{\hat{\rho} - \rho_A}  - 1\right),\,
\mu_B\left(\frac{r_{\ast\ast}}{r_{\ast}}\cdot\frac{\rho_A}{\hat{\rho} - \rho_B}-1\right)
\right\}.
\]
The (unique) interval of exchange ratios on which all shareholders agree on corresponds to a point on the curve $\delta$, represented by
\[
\left(\frac{\underline r_{\rho}(\hat{\rho}) + \overline r_{\rho}(\hat{\rho})}{2},\;
\frac{\overline r_{\rho}(\hat{\rho}) -  \underline r_{\rho}(\hat{\rho})}{2}\right),
\]
and corresponds to the region 
$\mathcal{BR}_{\rho}\left(\hat{\rho}\right)$, which shrinks to a singleton, in Table \ref{Table01}.

\begin{figure}
\centering
\includegraphics[width=.55\linewidth]{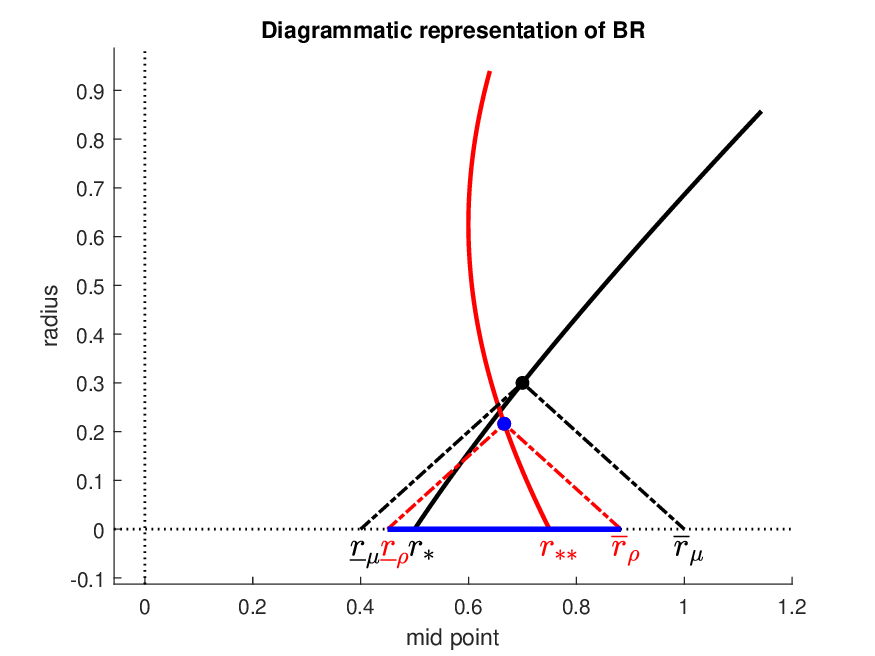}
\caption{Kulpa's diagrammatic representation of set $\mathcal{BR}_{\mu,\rho}$ - $p_A = 4$, $p_B = 2$, $\varphi_A = 4$, $\varphi_b = 3$, $N_A = 20$, $N_B = 10$, $r_{\ast}=0.5$, $r_{\ast\ast}=0.6667$, $\mu_M = 120$, $\rho_M = 98$. The resulting interval $\mathcal{BR}_{\rho}$ is denoted with the blue dot and segment.} \label{Figura-5c}
\end{figure}

The diagrammatic representation for this case is in Figure \ref{Figura-5c}.

\noindent \textbf{Case 4: Consistency region} $\mathcal{CR}_{A}\left(\mu_M,\hat{\rho}\right)$

\noindent Assume that $\underline r_{\mu}(\mu_M)\leq\underline r_{\rho}(\hat{\rho})$ and $\overline r_{\mu}(\mu_M)\leq \overline r_{\rho}(\hat{\rho})$. This situation is consistent, together with $\mu_M\geq \mu_A + \mu_B$, with the condition
\[
\max\left\{\dfrac{r_{\ast}}{r_{\ast\ast}}\cdot\dfrac{\mu_A\rho_B}{\hat{\rho} - \rho_A} + \mu_B,\,\dfrac{r_{\ast\ast}}{r_{\ast}}\cdot\dfrac{\mu_B(\hat{\rho} - \rho_A)}{\rho_B} + \mu_A\right\}\leq \mu_M\leq\dfrac{r_{\ast\ast}}{r_{\ast}}\cdot\dfrac{\mu_B\rho_A}{\hat{\rho} - \rho_B} + \mu_A.
\]
The same condition, written in terms of expected synergy, becomes
\[
\begin{aligned}
\max &\left\{\mu_A\left(\dfrac{r_{\ast}}{r_{\ast\ast}}\cdot\dfrac{\rho_B}{\hat{\rho} - \rho_A} -  1\right),\,\mu_B\left(\dfrac{r_{\ast\ast}}{r_{\ast}}\cdot\dfrac{\hat{\rho} - \rho_A}
{\rho_B} - 1\right)\right\}\leq \mu_M - \mu_A - \mu_B \\
&\leq\mu_B\left(\dfrac{r_{\ast\ast}}{r_{\ast}}\cdot\dfrac{\rho_A}{\hat{\rho} - \rho_B} - 1\right).
\end{aligned}
\]
The set of exchange ratios on which all shareholders agree, represented, according to Kulpa, by the points
\[
\left( \frac{\underline r_{\rho}(\hat{\rho}) + \overline r_{\mu}(\mu_M)}{2},\, \frac{ \overline r_{\mu}(\mu_M) - \underline r_{\rho}(\hat{\rho})}{2}\right),
\]
lies on the straight line of equation
\[
y = x - \frac{r_{\ast\ast}(\hat{\rho} - \rho_A)}{\rho_B},
\]
which corresponds to the region  $\mathcal{CR}_{A}\left(\mu_M,\hat{\rho}\right)$ in Table \ref{Table01}.

If we fix instead $\hat{\mu}\geq\mu_A + \mu_B$, we can consider the point $P_\mu(\hat{\mu})$, corresponding to the bargaining region $\mathcal{BR}_\mu(\hat{\mu})$.  
Recall that the intersection of the sets is not empty if \eqref{condizione_per_intersezione} holds, then, again, this represents the starting point for the analysis of the possible scenarios.  The details can be found in Appendix \ref{App-4}.

\begin{figure}
\centering
\includegraphics[width=.55\linewidth]{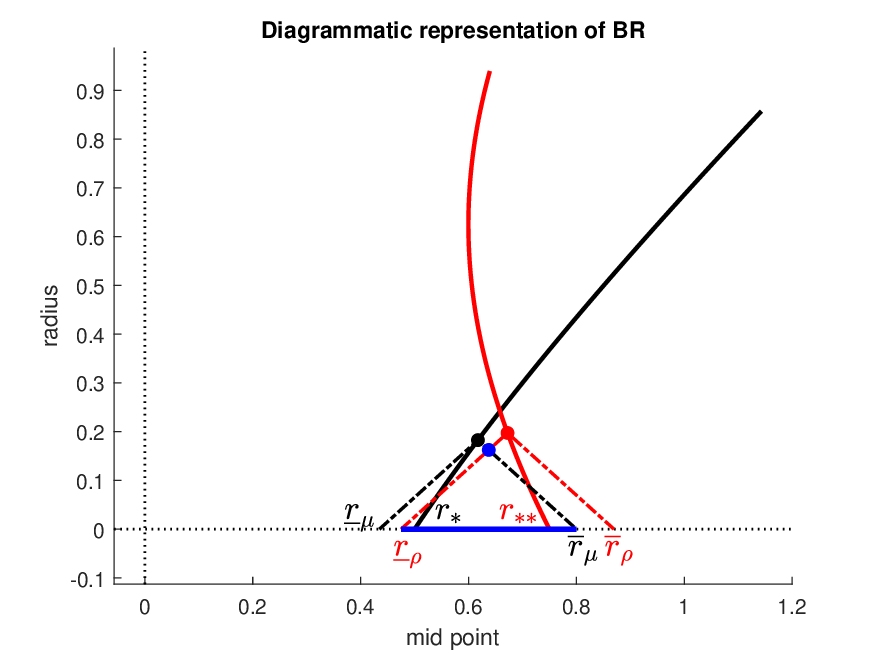}
\caption{Kulpa's diagrammatic representation of set $\mathcal{BR}_{\mu,\rho}$ - $p_A = 4$, $p_B = 2$, $\varphi_A = 4$, $\varphi_b = 3$, $N_A = 20$, $N_B = 10$, $r_{\ast}=0.5$, $r_{\ast\ast}=0.6667$, $\mu_M = 112$, $\rho_M = 99$. The resulting interval $\mathcal{CR}_{A}$ is denoted with the blue dot and segment.} \label{Figura-5d}
\end{figure}

The diagrammatic representation for this case is in Figure \ref{Figura-5d}.

\noindent \textbf{Case 1: Consistency region} $\mathcal{CR}_{B}\left(\hat{\mu},\rho_M\right)$

\noindent  Assume that $\underline r_{\mu}(\hat{\mu})\geq\underline r_{\rho}(\rho_M)$ and $\overline r_{\mu}(\hat{\mu})\geq \overline r_{\rho}(\rho_M)$. This situation is consistent, together with $\max\{\rho_A,\rho_B\}<\rho_M\leq \rho_A + \rho_B$, with the condition
$$
\frac{r_{\ast\ast}}{r_\ast}\cdot\frac{\mu_B\rho_A}{\hat{\mu}-\mu_A}+\rho_B\leq\rho_M\leq \min\left\{\frac{r_\ast}{r_{\ast\ast}}\cdot\frac{\mu_A\rho_B}{\hat{\mu}-\mu_B}+\rho_A,
\frac{r_{\ast\ast}}{r_\ast}\cdot\frac{(\hat{\mu}-\mu_B)\rho_A}{\mu_A}+\rho_B\right\}.
$$
The same condition, written in terms of reduction of the overall risk, becomes
$$
\begin{aligned}
\max &\left\{\rho_B\left(1 -\frac{r_\ast}{r_{\ast\ast}}\cdot\frac{\mu_A}{\hat{\mu}-\mu_B}\right),\,\rho_A\left(1 -\frac{r_{\ast\ast}}{r_\ast}\cdot\frac{\hat{\mu}-\mu_B}{\mu_A}\right)
\right\}\leq \rho_A+\rho_B-\rho_M \\
&\leq \rho_A\left(1 -\frac{r_{\ast\ast}}{r_\ast}\cdot\frac{\mu_B}{\hat{\mu}-\mu_A}\right).
\end{aligned}
$$
The set of exchange ratios on which all shareholders agree,  represented, according to Kulpa, by the points
\[
\left( \frac{\underline r_{\mu}(\hat{\mu}) + \overline r_{\rho}(\rho_M)}{2},\, \frac{ \overline r_{\rho}(\rho_M)- \underline r_{\mu}(\hat{\mu})}{2}\right),
\]
lies on the straight line of equation
\[
y = x - \frac{r_{\ast}\mu_A}{\hat{\mu} - \mu_B},
\]
which corresponds to the region  $\mathcal{CR}_{B}\left(\hat{\mu},\rho_M\right)$ in Table \ref{Table01}.

\noindent \textbf{Case 2: Bargaining  region} $\mathcal{BR}_{\mu}\left(\hat{\mu}\right)$

\noindent Assume that $\underline r_{\mu}(\hat{\mu})\geq\underline r_{\rho}(\rho_M)$ and $\overline r_{\mu}(\hat{\mu})\leq \overline r_{\rho}(\rho_M)$. This situation is consistent, together with $\max\{\rho_A,\rho_B\}<\rho_M\leq \rho_A + \rho_B$, with the condition
\[
\max\{\rho_A,\rho_B\}\leq \rho_M \leq\min\left\{\dfrac{r_{\ast}}{r_{\ast\ast}}\cdot\dfrac{\rho_B\mu_A}{\hat{\mu} - \mu_B} + \rho_A,\,\dfrac{r_{\ast\ast}}{r_{\ast}}\cdot\dfrac{\rho_A\mu_B}{\hat{\mu} - \mu_A} + \rho_B\right\}.
\]
The same condition, written in terms of overall risk, becomes
\[
\max\left\{\rho_B\left(1 - \dfrac{r_{\ast}}{r_{\ast\ast}}\cdot\dfrac{\rho_B\mu_A}{\hat{\mu} - \mu_B} \right),\,\rho_A\left(1 - \dfrac{r_{\ast\ast}}{r_{\ast}}\cdot\dfrac{\rho_A\mu_B}{\hat{\mu} - \mu_A} \right)\right\}\leq\rho_A + \rho_B - \rho_M <\min\{\rho_A,\rho_B\}.
\]
The (unique) interval of exchange ratios on which all shareholders agree on corresponds to the point on the curve $\gamma$
\[
\left(\frac{\underline r_{\mu}(\hat{\mu}) + \overline r_{\mu}(\hat{\mu})}{2},\;
\frac{\overline r_{\mu}(\hat{\mu}) -  \underline r_{\mu}(\hat{\mu})}{2}\right),
\]
representing the region $\mathcal{BR}_{\mu}\left(\hat{\mu}\right)$ in Table \ref{Table01}.



\noindent \textbf{Case 3: Bargaining region} $\mathcal{BR}_{\rho}\left(\rho_M\right)$

\noindent Assume that $\underline r_{\mu}(\hat{\mu})\leq\underline r_{\rho}(\rho_M)$ and $\overline r_{\mu}(\hat{\mu})\geq \overline r_{\rho}(\rho_M)$. This situation is consistent, together with $\max\{\rho_A,\rho_B\}<\rho_M\leq \rho_A + \rho_B$, with the condition
\[
\rho_M \geq\max\left\{\dfrac{r_{\ast}}{r_{\ast\ast}}\cdot\dfrac{\rho_B\mu_A}{\hat{\mu} - \mu_B} + \rho_A,\,\dfrac{r_{\ast\ast}}{r_{\ast}}\cdot\dfrac{\rho_A\mu_B}{\hat{\mu} - \mu_A} + \rho_B\right\}.
\]
The same condition, written in terms of overall risk, becomes
\[
\rho_A + \rho_B - \rho_M \leq\min\left\{\rho_B\left(1 - \dfrac{r_{\ast}}{r_{\ast\ast}}\cdot\dfrac{\rho_B\mu_A}{\hat{\mu} - \mu_B}\right) ,\,\rho_A\left(1 - \dfrac{r_{\ast\ast}}{r_{\ast}}\cdot\dfrac{\rho_A\mu_B}{\hat{\mu} - \mu_A}\right) \right\}.
\]
The set of exchange ratios on which all shareholders agree on lies on the curve $\delta$, and corresponds to the region 
$\mathcal{BR}_{\rho}\left(\rho_M\right)$ in Table \ref{Table01}.

\noindent \textbf{Case 4: Consistency region} $\mathcal{CR}_{A}\left(\hat{\mu},\rho_M\right)$

\noindent  Assume that $\underline r_{\mu}(\hat{\mu})\leq\underline r_{\rho}(\rho_M)$ and $\overline r_{\mu}(\hat{\mu})\leq \overline r_{\rho}(\rho_M)$. This situation is consistent, together with $\max\{\rho_A,\rho_B\}<\rho_M\leq \rho_A + \rho_B$, with the condition
$$
\frac{r_\ast}{r_{\ast\ast}}\cdot\frac{\rho_B\mu_A}{\hat{\mu}-\mu_B}+\rho_A\leq\rho_M\leq \min\left\{\frac{r_{\ast\ast}}{r_\ast}\cdot\frac{\rho_A\mu_B}{\hat{\mu}-\mu_A}+\rho_B,
\frac{r_\ast}{r_{\ast\ast}}\cdot\frac{\rho_B(\hat{\mu}-\mu_A)}{\mu_B}+\rho_A\right\}.
$$
The same condition, written in terms of reduction of the overall risk, becomes
$$
\begin{aligned}
\max &\left\{\rho_A\left(1 -\frac{r_{\ast\ast}}{r_\ast}\cdot\frac{\mu_B}{\hat{\mu}-\mu_A}\right),\,\rho_B\left(1 -\frac{r_\ast}{r_{\ast\ast}}\cdot\frac{\hat{\mu}-\mu_A}{\mu_B}\right)
\right\}\leq \rho_A+\rho_B-\rho_M \\
&\leq \rho_B\left(1 -\frac{r_\ast}{r_{\ast\ast}}\cdot\frac{\mu_A}{\hat{\mu}-\mu_B}\right).
\end{aligned}
$$
The set of exchange ratios on which all shareholders agree,  represented, according to Kulpa, by the points
\[
\left( \frac{\underline r_{\rho}(\rho_M) + \overline r_{\mu}(\hat{\mu})}{2},\, \frac{ \overline r_{\mu}(\hat{\mu}) - \underline r_{\rho}(\rho_M)}{2}\right),
\]
lies on the straight line of equation
\[
y = - x + \frac{r_{\ast}(\hat{\mu} - \mu_A)}{\mu_B},
\]
which corresponds to the region  $\mathcal{CR}_{A}\left(\hat{\mu},\rho_M\right)$ in Table \ref{Table01}.

\begin{figure}
\centering
\includegraphics[width=.55\linewidth]{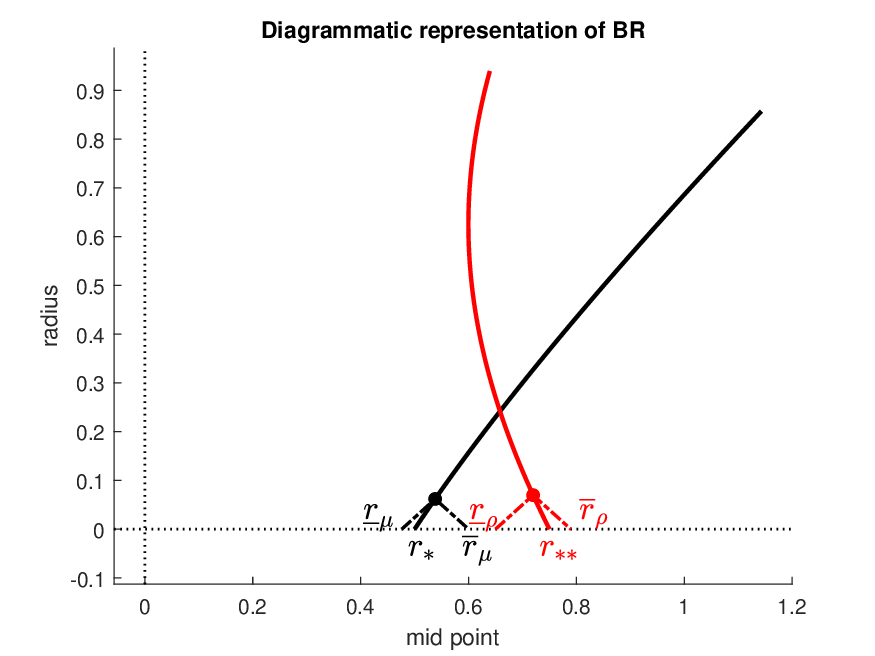}
\caption{Kulpa's diagrammatic representation of set $\mathcal{BR}_{\mu,\rho}$ - $p_A = 4$, $p_B = 2$, $\varphi_A = 4$, $\varphi_b = 3$, $N_A = 20$, $N_B = 10$, $r_{\ast}=0.5$, $r_{\ast\ast}=0.6667$, $\mu_M = 104$, $\rho_M = 106$. The resulting set is empty.} \label{Figura-5e}
\end{figure}

Two more cases are displayed in this Table: in both, sets $\mathcal{BR}_{\mu,\rho}\left(\mu,\rho\right)$ is empty. Figure \ref{Figura-5e} shows one of them. As it is clear from the above presentation, this set is empty due to the limited increase in expected equity value and/or limited reduction in overall risk.

The last case resembles the one just presented, with the only difference that $r_{\ast\ast} < r_{\ast}$. For sake of paucity, its diagrammatic representation is omitted.

\section{Conclusions}\label{Conclusions}

In this article, we present an attempt to extend, in a stochastic setting, deterministic models for exchange ratios determination for merger agreements.

Under a financial point of view, an important result we achieve is that the introduction of a risk measure changes the attitude shareholders of the merging companies have toward the agreed exchange ratio. For shareholders of the acquiring company, smaller exchange ratios result in both larger expected equity value and equity risk. For shareholders of the acquired company, larger exchange ratio yield both a larger expected equity value and equity risk. This introduces a trade-off that makes the negotiation for settling on the definitive exchange ratio more fitting to a risky environment. 


As far as the bargaining region is concerned, being here  dependent on two quantities, its effective representation is obtained exploiting the Kulpa's diagrammatic technique as it transforms bounded intervals into points in a bi-dimensional plane and allows to plainly represent all possible cases.  

Even if it can be argued that, in real financial markets, imposing that a merger creates both a positive expected synergy and a reduction in the overall equity value risk rarely occurs, our contribution can be intended as a theoretical framework under which mergers can be analyzed, a result that lacks in financial literature.

To conclude, results presented here must be intended as a first step further research development in this field. For instance, ranges of acceptable exchange ratios might be determined following an expected utility approach, where a trade-off between expected value and risk of an investor's wealth can be established. 

\section*{Acknowledgments} The idea of exploiting the diagrammatic approach was suggested by Nicolae Popovici, wonderful friend, excellent researcher, and highly esteemed professor at the Department of Mathematics of the Faculty of Mathematics and Computer Science, Babeș-Bolyai University in Cluj-Napoca, Romania. Sadly, Nicolae suddenly passed away in June 2022. Without Nicolae's help, this article would lack its best part.

La revedere, Nicolae; și mulțumesc pentru tot!


\appendix

\section{Appendices}

\subsection{Appendix 1} \label{App1}

To justify the claim about Company $B$'s stockholders in subsection \ref{SubMeAr-2}, consider at first what happens in the left-hand side of (\ref{Cond2SDbis}) if $r=0$: company $B$'s shareholders receive no stock, carry no risk, and experience a reduction in their riskiness. 

As
$$
\frac{\partial}{\partial r} \left[ \frac{r\cdot\rho_M}{N_A + rN_B}
\right] = \frac{\rho_M N_A}{(N_A + r N_B)^2} >0\text{,}
$$
the left-hand side of expression (\ref{Cond2SDbis}) is strictly increasing with respect to $r \ge 0$ so that the larger $r$, the larger the risk for $B$'s stockholders. 

Finally, being, in accordance with (\ref{range-rho}), $\rho_M > \rho_B$,
$$
\lim_{r \to +\infty} \frac{r\cdot\rho_M}{N_A + rN_B} = \frac{\rho_M}{N_B} > \frac{\rho_B}{N_B}\text{.}
$$
Then, it results that sufficiently large values for $r$ bring about larger risk to company $B$'s stockholders when compared to the risk level they bore before companies  merged.

\subsection{Appendix 2} \label{App2}


Consider point $P_{\mu}(\mu_M)$ in (\ref{punto_iperbole_mu}). The difference between its coordinates is
\begin{equation}
x(\mu_M) - y(\mu_M) = \frac{r_{\ast}\mu_A}{\mu_M - \mu_B}, 
\end{equation}
from which (removing the dependence from $\mu_M$ for the sake of simplicity) we get
\[
\mu_M = \frac{r_{\ast}\mu_A}{x - y} + \mu_B.
\]
Recalling that
\begin{align*}
x &= \frac{r_{\ast}}{2}\left(\frac{\mu_A}{\mu_M - \mu_B} + \frac{\mu_M - \mu_A}{\mu_B}\right) \\
y &= \frac{r_\ast}{2}\left(\frac{\mu_M-\mu_A}{\mu_B}-\frac{\mu_A}{\mu_M-\mu_B}\right)
\end{align*}
and replacing the expression of $\mu_M$ found above, we obtain
$$
x+y=r_\ast\left(\frac{r_\ast\mu_A}{\mu_B}\cdot\frac{1}{x-y}+1-\frac{\mu_A}{\mu_B}\right)
$$
and multiplying by $x-y$ the following equation is obtained
$$
x^2-y^2=r_\ast\left(1-\frac{\mu_A}{\mu_B}\right)(x-y)+r_\ast^2\frac{\mu_A}{\mu_B},
$$
so that $\gamma$ is given by
\begin{equation}
\label{gamma}
x^2-y^2-r_\ast\left(1-\frac{\mu_A}{\mu_B}\right)(x-y)-r_\ast^2\frac{\mu_A}{\mu_B}=0.
\end{equation}
This is the equation of the quadratic curve $$c_1 x^2 + c_2 xy + c_3 y^2 + c_4 x + c_5y + c_6 = 0$$ with $c_1 = 1$, $c_2 = 0$, $c_3 = -1$, $c_4 = -r_{\ast} \left(\mu_B - \mu_A\right)/\mu_B$, $c_5 = r_{\ast} \left(\mu_B - \mu_A\right)/\mu_B$, and $c_6=-r_{\ast}^2\mu_A/\mu_B$.

As $c_2^2-4c_1 c_3 =4>0$, equation \eqref{gamma} identifies a hyperbola whose eccentricity is $\sqrt{2}$ and that 
	can be rewritten as
	\begin{equation}
		\frac{\left(x - r_{\ast} \frac{\mu_B - \mu_A}{2\mu_B}\right)^2}{r^2_{\ast}\frac{\mu_A}{\mu_B}} - 
		\frac{\left(y - r_{\ast} \frac{\mu_B - \mu_A}{2\mu_B}\right)^2}{r^2_{\ast}\frac{\mu_A}{\mu_B}} = 1 \label{Iperbole-gamma}
	\end{equation}
	
	
	
	Its two vertices are
	\begin{equation*}
		V_1 \equiv \left(\frac{r_{\ast}\left(\mu_B - \mu_A\right)}{2\mu_B} - r_{\ast} \sqrt{\frac{\mu_A}{\mu_B}},\frac{r_{\ast}\left(\mu_B - \mu_A\right)}{2\mu_B}\right)\text{,}
	\end{equation*}
	and 
	\begin{equation*}
		V_2 \equiv \left(\frac{r_{\ast}\left(\mu_B - \mu_A\right)}{2\mu_B} + r_{\ast} \sqrt{\frac{\mu_A}{\mu_B}},\frac{r_{\ast}\left(\mu_B - \mu_A\right)}{2\mu_B}\right)
	\end{equation*}
	
	
	The abscissa of $V_1$ is smaller then the abscissa of $V_2$. In terms of diagrammatic representation (see Figure \ref{Figura-2a}), the relevant branch of hyperbola \eqref{Iperbole-gamma} is the one whose vertex is $V_2$. This is the case as there exist points on the hyperbola's branch that contains $V_1$ with negative abscissas, a condition that is not compatible with the fact that points belonging to $P_{\mu}\left(\mu_M\right)$ must have positive $x-$values. 
	
	Focusing on $V_2$, its abscissa is surely positive if $\mu_B > \mu_A$. If, instead, $\mu_B < \mu_A$ then the abscissa of $V_2$ is positive as long as $\mu_B \ge \left(3 - 2\sqrt{2} \right) \mu_A$.
	
	
	As far as the ordinate of $V_2$ is concerned, it is positive (respectively negative) when $\mu_B > \mu_A$ $\left(\text{respectively }\mu_B < \mu_A\right)$. 
	
	The asymptotes of \eqref{Iperbole-gamma} are
	$$
	y=x, \qquad y=-x+  \frac{r_\ast\left(\mu_B-\mu_A\right)}{\mu_B}.
	$$
	
	Similarly, consider point $P_{\rho}(\rho_M)$ in \eqref{punto_iperbole_rho}. The sum of its coordinates is
	$$
	x(\rho_M) + y(\rho_M) = r_{\ast\ast}\frac{\rho_A}{\rho_M-\rho_B}, 
	$$
	from which (removing the dependence from $\rho_M$ for the sake of simplicity) we get
	\[
	\rho_M = \frac{r_{\ast\ast}\rho_A}{x+y} + \rho_B.
	\]
	Recalling that
	\begin{align*}
		x &= \frac{r_{\ast\ast}}{2}\left(\frac{\rho_M-\rho_A}{\rho_B} + \frac{\rho_A}{\rho_M-\rho_B}\right) \\
		y &= \frac{r_{\ast\ast}}{2}\left(\frac{\rho_A}{\rho_M-\rho_B}-\frac{\rho_M-\rho_A}{\rho_B}\right)
	\end{align*}
	and replacing the expression of $\rho_M$ found above, we obtain
	$$
	x-y=r_{\ast\ast}\left(\frac{r_{\ast\ast}\rho_A}{\rho_B}\cdot\frac{1}{x+y}+1-\frac{\rho_A}{\rho_B}\right)
	$$
	and multiplying by $x+y$ the following equation is obtained
	$$
	x^2-y^2=r_{\ast\ast}\left(1-\frac{\rho_A}{\rho_B}\right)(x+y)+r_{\ast\ast}^2\frac{\rho_A}{\rho_B},
	$$
	so that $\delta$ is given by
	\begin{equation}
		\label{delta}
		x^2-y^2-r_{\ast\ast}\left(1-\frac{\rho_A}{\rho_B}\right)(x+y)-r_{\ast\ast}^2\frac{\rho_A}{\rho_B}=0,
	\end{equation}
	which is, again, a hyperbola with equation
	$$
	\frac{\left(x - r_{\ast\ast} \frac{\rho_B - \rho_A}{2\rho_B}\right)^2}{r^2_{\ast\ast} \frac{\rho_A}{\rho_B}} - \frac{\left(y + 
		r_{\ast\ast} \frac{\rho_B - \rho_A}{2\rho_B}\right)^2}{r^2_{\ast\ast} \frac{\rho_A}{\rho_B}} = 1.
	$$
	
	Its vertices are
	$$
	V_3 \equiv\left(\frac{r_{\ast\ast}\left(\rho_B - \rho_A\right)}{2\rho_B} - r_{\ast\ast} \sqrt{\frac{\rho_A}{\rho_B}},\frac{r_{\ast\ast}\left(\rho_B - \rho_A\right)}{2\rho_B}\right)
	$$
	and 
	$$
	V_4 \equiv\left(\frac{r_{\ast\ast}\left(\rho_B - \rho_A\right)}{2\rho_B} + r_{\ast\ast} \sqrt{\frac{\rho_A}{\rho_B}},\frac{r_{\ast
			\ast}\left(\rho_B - \rho_A\right)}{2\rho_B}\right)
	$$
while the asymptotes are
$$
y = -x, \quad  
y = x - \frac{r_{\ast\ast}\left(\rho_B - \rho_A\right)}{\rho_B}.
$$
The abscissa of $V_3$ is smaller then the abscissa of $V_4$. In terms of diagrammatic representation (see Figure \ref{Figura-2a}), the relevant branch of hyperbola \eqref{delta} is the one whose vertex is $V_4$. This is the case as there exist points on the hyperbola's branch that contains $V_3$ with negative abscissas, a condition that is not compatible with the fact that points belonging to $P_{\rho}\left(\rho_M\right)$ must have positive $x-$values. 

Focusing on $V_4$, its abscissa is surely positive if $\rho_B > \rho_A$. If, instead, $\rho_B < \rho_A$ then the abscissa of $V_4$ is positive as long as $\rho_B \ge \left(3 - 2\sqrt{2} \right) \rho_A$.

\subsection{Appendix 3} \label{App-3}
In terms of risk-adjusted performance, the resulting company performs better then the pre-existing ones when
$$
\lambda_M = \frac{\mu_A + \mu_B + s}{\rho_A + \rho_B - v}
\ge \frac{\mu_{A}}{\rho_{A}} = \lambda_A\text{,}
$$
and
$$
\lambda_M = \frac{\mu_A + \mu_B + s}{\rho_A + \rho_B - v}
\ge \frac{\mu_{B}}{\rho_{B}} = \lambda_B\text{.}
$$
Recalling the notation for expected synergy and risk reduction introduced in Subsection (\ref{SubMeAr-2}), these inequalities are respectively equivalent to
$$
\frac{\mu_B + s}{\rho_B -v} \ge \lambda_A\text{,}
$$
and
$$
\frac{\mu_A + s}{\rho_A -v} \ge \lambda_B\text{.}
$$
If, now, $s = 0$ and $v = 0$, the first inequality becomes
$$
\frac{\mu_B}{\rho_B} = \lambda_B \ge \lambda_A
$$
while the second reads
$$
\frac{\mu_A}{\rho_A} = \lambda_A \ge \lambda_B\text{.}
$$
These expressions explain why, in the absence of any improvement in terms of either positive expected synergy or risk reduction, stockholders of company $A$ $\left(\text{respectively }B\right)$ experience an improvement in the risk-corrected performance of the shares they own if company $B$ $\left(\text{respectively }A\right)$ better behaves, in terms of risk-corrected performance.

\subsection{Appendix 4} \label{App-4}
Consider $\max\{\rho_A,\rho_B\} < \hat{\rho}\leq \rho_A + \rho_B$. In Case 1, the intersection is the set $\mathcal{CR}_B\left(\mu_M,\hat{\rho}\right) = \left[\underline r_{\mu}(\mu_M),
\overline r_{\rho}(\hat{\rho})\right]$. This interval is represented, under Kulpa's approach, by the point
\[
(x,y)  = \left(\frac{\underline r_{\mu}(\mu_M) + \overline r_{\rho}(\hat{\rho})}{2},\,\frac{\overline r_{\rho}(\hat{\rho}) - \underline r_{\mu}(\mu_M)}{2}\right),
\]
that is,
\[
x = \frac{1}{2}\left(\frac{r_{\ast}\mu_A}{\mu_M - \mu_B} + \frac{r_{\ast\ast}\rho_A}{\hat{\rho} - \rho_B}\right),
\]
\[
y = \frac{1}{2}\left(\frac{r_{\ast\ast}\rho_A}{\hat{\rho} - \rho_B} - \frac{r_{\ast}\mu_A}{\mu_M - \mu_B}\right).
\]
The sum of its coordinates is
\[
x + y = \frac{r_{\ast\ast}\rho_A}{\hat{\rho} - \rho_B};
\]
that is, on varying $\mu_M$ we move along the straight line of equation
\[
y = - x + \frac{r_{\ast\ast}\rho_A}{\hat{\rho} - \rho_B}.
\]
The conditions under which this situation is possible are
\[
\begin{sistema}
	\smallskip\underline r_{\mu}(\mu_M)\geq \underline r_{\rho}(\hat{\rho})\\
	\smallskip\overline r_{\mu}(\mu_M)\geq \overline r_{\rho}(\hat{\rho})\\
	\underline r_{\mu}(\mu_M)\leq \overline r_{\rho}(\hat{\rho}),
\end{sistema}
\]
that is,
\[
\begin{sistema}
	\medskip
	\dfrac{r_{\ast}\mu_A}{\mu_M - \mu_B}\geq \dfrac{r_{\ast\ast}(\hat{\rho} - \rho_A)}{\rho_B}\\
	\medskip
	\dfrac{r_{\ast}(\mu_M - \mu_A)}{\mu_B}\geq \dfrac{r_{\ast\ast}\rho_A}{\hat{\rho} - \rho_B}\\
	\dfrac{r_{\ast}\mu_A}{\mu_M - \mu_B}\leq\dfrac{r_{\ast\ast}\rho_A}{\hat{\rho} - \rho_B},
\end{sistema}
\;\;\;\;\;\text{that is,}\;\;\;\;\;
\begin{sistema}
	\medskip 
	\mu_M \leq\dfrac{r_{\ast}}{r_{\ast\ast}}\cdot\dfrac{\mu_A\rho_B}{\hat{\rho} - \rho_A} + \mu_B\\
	\medskip
	\mu_M \geq \dfrac{r_{\ast\ast}}{r_{\ast}}\cdot\dfrac{\mu_B\rho_A}{\hat{\rho} - \rho_B} + \mu_A\\
	\mu_M \geq \dfrac{r_{\ast}}{r_{\ast\ast}}\cdot\dfrac{\mu_A(\hat{\rho} - \rho_B)}{\rho_A} + \mu_B,
\end{sistema}
\]
which can be summarized as follows:
\[
\max\left\{\dfrac{r_{\ast\ast}}{r_{\ast}}\cdot\dfrac{\mu_B\rho_A}{\hat{\rho} - \rho_B} + \mu_A,\,\dfrac{r_{\ast}}{r_{\ast\ast}}\cdot\dfrac{\mu_A(\hat{\rho} - \rho_B)}{\rho_A} + \mu_B\right\}\leq \mu_M \leq\dfrac{r_{\ast}}{r_{\ast\ast}}\cdot\dfrac{\mu_A\rho_B}{\hat{\rho} - \rho_A} + \mu_B.
\]
To obtain the condition about the expected synergy, we subtract $\mu_A + \mu_B$:
\[
\max\left\{\dfrac{r_{\ast\ast}}{r_{\ast}}\cdot\dfrac{\mu_B\rho_A}{\hat{\rho} - \rho_B} - \mu_B,\,\dfrac{r_{\ast}}{r_{\ast\ast}}\cdot\dfrac{\mu_A(\hat{\rho} - \rho_B)}{\rho_A} - \mu_A\right\}\leq \mu_M - \mu_A - \mu_B \leq\dfrac{r_{\ast}}{r_{\ast\ast}}\cdot\dfrac{\mu_A\rho_B}{\hat{\rho} - \rho_A} - \mu_A,
\]
that is,
\[
\begin{aligned}
	\max &\left\{\mu_B\left(\dfrac{r_{\ast\ast}}{r_{\ast}}\cdot\dfrac{\rho_A}{\hat{\rho} - \rho_B} - 1\right),\,\mu_A\left(\dfrac{r_{\ast}}{r_{\ast\ast}}\cdot\dfrac{\hat{\rho} - \rho_B}
	{\rho_A} - 1\right)\right\}\leq \mu_M - \mu_A - \mu_B \\
	&\leq\mu_A\left(\dfrac{r_{\ast}}{r_{\ast\ast}}\cdot\dfrac{\mu_A\rho_B}{\hat{\rho} - \rho_A} - 1\right).
\end{aligned}
\]
In Case 2, the intersection is the interval $\mathcal{BR}_{\mu}\left(\mu_M\right) = \left[\underline r_{\mu}(\mu_M),\overline r_{\mu}(\mu_M)\right]$. This is represented, under Kulpa's approach, by the point
\[
(x,y)  = \left(\frac{\underline r_{\mu}(\mu_M) + \overline r_{\mu}(\mu_M)}{2},\,\frac{\overline r_{\mu}(\mu_M) - \underline r_{\mu}(\mu_M)}{2}\right),
\]
which, varying $\mu_M$, leads to the curve $\gamma$ (see \ref{App2}). The conditions under which this situation is possible are
\[
\begin{sistema}
	\smallskip\underline r_{\mu}(\mu_M)\geq \underline r_{\rho}(\hat{\rho})\\
	\overline r_{\mu}(\mu_M)\leq \overline r_{\rho}(\hat{\rho}),
\end{sistema}
\]
that is,
\[
\begin{sistema}
	\medskip
	\dfrac{r_{\ast}\mu_A}{\mu_M - \mu_B}\geq \dfrac{r_{\ast\ast}(\hat{\rho} - \rho_A)}{\rho_B}\\
	\dfrac{r_{\ast}(\mu_M - \mu_A)}{\mu_B}\leq \dfrac{r_{\ast\ast}\rho_A}{\hat{\rho} - \rho_B},
\end{sistema}
\;\;\;\;\;\text{that is,}\;\;\;\;\;
\begin{sistema}
	\medskip 
	\mu_M \leq\dfrac{r_{\ast}}{r_{\ast\ast}}\cdot\dfrac{\mu_A\rho_B}{\hat{\rho} - \rho_A} + \mu_B\\
	\mu_M \leq \dfrac{r_{\ast\ast}}{r_{\ast}}\cdot\dfrac{\mu_B\rho_A}{\hat{\rho} - \rho_B} + \mu_A,
\end{sistema}
\]
which can be summarized as follows:
\[
\mu_A + \mu_B\leq \mu_M \leq\min\left\{\dfrac{r_{\ast}}{r_{\ast\ast}}\cdot\dfrac{\mu_A\rho_B}{\hat{\rho} - \rho_A} + \mu_B,\,\dfrac{r_{\ast\ast}}{r_{\ast}}\cdot\dfrac{\mu_B\rho_A}{\hat{\rho} - \rho_B} + \mu_A\right\}.
\]
To obtain the condition about the expected synergy, we subtract $\mu_A + \mu_B$:
\[
0 \leq \mu_M - \mu_A - \mu_B\leq\min\left\{\dfrac{r_{\ast}}{r_{\ast\ast}}\cdot\dfrac{\mu_A\rho_B}{\hat{\rho} - \rho_A} - \mu_A,\,\dfrac{r_{\ast\ast}}{r_{\ast}}\cdot\dfrac{\mu_B\rho_A}{\hat{\rho} - \rho_B} - \mu_B\right\},
\]
that is,
\[
0 \leq \mu_M - \mu_A - \mu_B\leq\min\left\{\mu_A\left(\dfrac{r_{\ast}}{r_{\ast\ast}}\cdot\dfrac{\rho_B}{\hat{\rho} - \rho_A} - 1\right),\,\mu_B\left(\dfrac{r_{\ast\ast}}{r_{\ast}}\cdot\dfrac{\rho_A}{\hat{\rho} - \rho_B} - 1\right)\right\}.
\]
In Case 3, the intersection is the interval $\mathcal{BR}_{\rho}\left(\hat{\rho}\right) = \left[\underline r_{\rho}(\hat{\rho}),\overline r_{\rho}(\hat{\rho})\right]$.  This is represented, under Kulpa's approach, by the point
\[
(x,y)  = \left(\frac{\underline r_{\rho}(\hat{\rho}) + \overline r_{\rho}(\hat{\rho})}{2},\,\frac{\overline r_{\rho}(\hat{\rho}) - \underline r_{\rho}(\hat{\rho})}{2}\right).
\]
This point is unique, that is, the bargaining region does not depend on $\mu_M$.  The conditions under which this situation is possible are
\[
\begin{sistema}
	\smallskip\underline r_{\mu}(\mu_M)\leq \underline r_{\rho}(\hat{\rho})\\
	\overline r_{\mu}(\mu_M)\geq \overline r_{\rho}(\hat{\rho}),
\end{sistema}
\]
that is,
\[
\begin{sistema}
	\medskip
	\dfrac{r_{\ast}\mu_A}{\mu_M - \mu_B}\leq \dfrac{r_{\ast\ast}(\hat{\rho} - \rho_A)}{\rho_B}\\
	\dfrac{r_{\ast}(\mu_M - \mu_A)}{\mu_B}\geq \dfrac{r_{\ast\ast}\rho_A}{\hat{\rho} - \rho_B},
\end{sistema}
\;\;\;\;\;\text{that is,}\;\;\;\;\;
\begin{sistema}
	\medskip 
	\mu_M \geq\dfrac{r_{\ast}}{r_{\ast\ast}}\cdot\dfrac{\mu_A\rho_B}{\hat{\rho} - \rho_A} + \mu_B\\
	\mu_M \geq \dfrac{r_{\ast\ast}}{r_{\ast}}\cdot\dfrac{\mu_B\rho_A}{\hat{\rho} - \rho_B} + \mu_A,
\end{sistema}
\]
which can be summarized as follows:
\[
\mu_M \geq\max\left\{\dfrac{r_{\ast}}{r_{\ast\ast}}\cdot\dfrac{\mu_A\rho_B}{\hat{\rho} - \rho_A} + \mu_B,\,\dfrac{r_{\ast\ast}}{r_{\ast}}\cdot\dfrac{\mu_B\rho_A}{\hat{\rho} - \rho_B} + \mu_A\right\}.
\]
To obtain the condition about the expected synergy, we subtract $\mu_A + \mu_B$:
\[
\mu_M - \mu_A - \mu_B \geq\max\left\{\dfrac{r_{\ast}}{r_{\ast\ast}}\cdot\dfrac{\mu_A\rho_B}{\hat{\rho} - \rho_A} - \mu_A,\,\dfrac{r_{\ast\ast}}{r_{\ast}}\cdot\dfrac{\mu_B\rho_A}{\hat{\rho} - \rho_B} - \mu_B\right\},
\]
that is,
\[
\mu_M - \mu_A - \mu_B \geq\max\left\{\mu_A\left(\dfrac{r_{\ast}}{r_{\ast\ast}}\cdot\dfrac{\rho_B}{\hat{\rho} - \rho_A} - 1\right),\,\mu_B\left(\dfrac{r_{\ast\ast}}{r_{\ast}}\cdot\dfrac{\rho_A}{\hat{\rho} - \rho_B} - 1\right)\right\}.
\]
In Case 4, the intersection is the interval $\mathcal{CR}_A\left(\mu_M,\hat{\rho}\right) = \left[\underline r_{\rho}(\hat{\rho}),\overline r_{\mu}(\mu_M)\right]$. This is represented, under Kulpa's approach, by the point
\[
(x,y)  = \left(\frac{\underline r_{\rho}(\hat{\rho}) + \overline r_{\mu}(\mu_M)}{2},\,\frac{\overline r_{\mu}(\mu_M) - \underline r_{\rho}(\hat{\rho})}{2}\right),
\]
that is,
\[
x = \frac{1}{2}\left(\frac{r_{\ast\ast}(\hat{\rho} - \rho_A)}{\rho_B} + \frac{r_{\ast}(\mu_M - \mu_A)}{\mu_B}\right),
\]
\[
y = \frac{1}{2}\left(\frac{r_{\ast}(\mu_M - \mu_A)}{\mu_B} - \frac{r_{\ast\ast}(\hat{\rho} - \rho_A)}{\rho_B}\right).
\]
The difference between its coordinates is
\[
x - y = \frac{r_{\ast\ast}(\hat{\rho} - \rho_A)}{\rho_B};
\]
that is, on varying $\mu_M$ we move along the straight line of equation
\[
y = x - \frac{r_{\ast\ast}(\hat{\rho} - \rho_A)}{\rho_B}.
\]
The conditions under which this situation is possible are
\[
\begin{sistema}
	\smallskip\underline r_{\mu}(\mu_M)\leq \underline r_{\rho}(\hat{\rho})\\
	\smallskip\overline r_{\mu}(\mu_M)\leq \overline r_{\rho}(\hat{\rho})\\
	\underline r_{\rho}(\hat{\rho})\leq \overline r_{\mu}(\mu_M),
\end{sistema}
\]
that is,
\[
\begin{sistema}
	\medskip
	\dfrac{r_{\ast}\mu_A}{\mu_M - \mu_B}\leq \dfrac{r_{\ast\ast}(\hat{\rho} - \rho_A)}{\rho_B}\\
	\medskip
	\dfrac{r_{\ast}(\mu_M - \mu_A)}{\mu_B}\leq \dfrac{r_{\ast\ast}\rho_A}{\hat{\rho} - \rho_B}\\
	\dfrac{r_{\ast\ast}(\hat{\rho} - \rho_A)}{\rho_B}\leq\dfrac{r_{\ast}(\mu_M - \mu_A)}{\mu_B},
\end{sistema}
\;\;\;\;\;\text{that is,}\;\;\;\;\;
\begin{sistema}
	\medskip 
	\mu_M \geq\dfrac{r_{\ast}}{r_{\ast\ast}}\cdot\dfrac{\mu_A\rho_B}{\hat{\rho} - \rho_A} + \mu_B\\
	\medskip
	\mu_M \leq \dfrac{r_{\ast\ast}}{r_{\ast}}\cdot\dfrac{\mu_B\rho_A}{\hat{\rho} - \rho_B} + \mu_A\\
	\mu_M \geq \dfrac{r_{\ast\ast}}{r_{\ast}}\cdot\dfrac{\mu_B(\hat{\rho} - \rho_A)}{\rho_B} + \mu_A,
\end{sistema}
\]
which can be summarized as follows:
\[
\max\left\{\dfrac{r_{\ast}}{r_{\ast\ast}}\cdot\dfrac{\mu_A\rho_B}{\hat{\rho} - \rho_A} + \mu_B,\,\dfrac{r_{\ast\ast}}{r_{\ast}}\cdot\dfrac{\mu_B(\hat{\rho} - \rho_A)}{\rho_B} + \mu_A\right\}\leq \mu_M\leq\dfrac{r_{\ast\ast}}{r_{\ast}}\cdot\dfrac{\mu_B\rho_A}{\hat{\rho} - \rho_B} + \mu_A.
\]
To obtain the condition about the expected synergy, we subtract $\mu_A + \mu_B$:
\[
\max\left\{\dfrac{r_{\ast}}{r_{\ast\ast}}\cdot\dfrac{\mu_A\rho_B}{\hat{\rho} - \rho_A} - \mu_A,\,\dfrac{r_{\ast\ast}}{r_{\ast}}\cdot\dfrac{\mu_B(\hat{\rho} - \rho_A)}{\rho_B} - \mu_B\right\}\leq \mu_M-\mu_A-\mu_B\leq\dfrac{r_{\ast\ast}}{r_{\ast}}\cdot\dfrac{\mu_B\rho_A}{\hat{\rho} - \rho_B} - \mu_B,
\]
that is,
\[
\begin{aligned}
	\max &\left\{\mu_A\left(\dfrac{r_{\ast}}{r_{\ast\ast}}\cdot\dfrac{\rho_B}{\hat{\rho} - \rho_A} -  1\right),\,\mu_B\left(\dfrac{r_{\ast\ast}}{r_{\ast}}\cdot\dfrac{\hat{\rho} - \rho_A}
	{\rho_B} - 1\right)\right\}\leq \mu_M - \mu_A - \mu_B \\
	&\leq\mu_B\left(\dfrac{r_{\ast\ast}}{r_{\ast}}\cdot\dfrac{\rho_A}{\hat{\rho} - \rho_B} - 1\right).
\end{aligned}
\]
Consider now $\hat{\mu}\geq \mu_A + \mu_B$. In Case 1, the intersection is the interval $\mathcal{CR}_B\left(\hat{\mu},\rho_M\right) = \left[\underline r_{\mu}(\hat{\mu}),\overline r_{\rho}(\rho_M)\right]$. This is represented, under Kulpa's approach, by the point
\[
(x,y)  = \left(\frac{\underline r_{\mu}(\hat{\mu}) + \overline r_{\rho}(\rho_M)}{2},\,\frac{\overline r_{\rho}(\rho_M) - \underline r_{\mu}(\hat{\mu})}{2}\right),
\]
that is,
\[
x = \frac{1}{2}\left(\frac{r_{\ast}\mu_A}{\hat{\mu} - \mu_B} + \frac{r_{\ast\ast}\rho_A}{\rho_M - \rho_B}\right),
\]
\[
y = \frac{1}{2}\left(\frac{r_{\ast\ast}\rho_A}{\rho_M - \rho_B} - \frac{r_{\ast}\mu_A}{\hat{\mu} - \mu_B}\right).
\]
The difference between its coordinates is
\[
x - y = \frac{r_{\ast}\mu_A}{\hat{\mu} - \mu_B};
\]
that is, on varying $\rho_M$ we move along the straight line of equation
\[
y = x - \frac{r_{\ast}\mu_A}{\hat{\mu} - \mu_B}.
\]
The conditions under which this situation is possible are
\[
\begin{sistema}
	\smallskip\underline r_{\mu}(\hat{\mu})\geq \underline r_{\rho}(\rho_M)\\
	\smallskip\overline r_{\mu}(\hat{\mu})\geq \overline r_{\rho}(\rho_M)\\
	\underline r_{\mu}(\hat{\mu})\leq \overline r_{\rho}(\rho_M),
\end{sistema}
\]
that is,
\[
\begin{sistema}
	\medskip
	\dfrac{r_{\ast}\mu_A}{\hat{\mu} - \mu_B}\geq \dfrac{r_{\ast\ast}(\rho_M - \rho_A)}{\rho_B}\\
	\medskip
	\dfrac{r_{\ast}(\hat{\mu} - \mu_A)}{\mu_B}\geq \dfrac{r_{\ast\ast}\rho_A}{\rho_M - \rho_B}\\
	\dfrac{r_{\ast}\mu_A}{\hat{\mu} - \mu_B}\leq\dfrac{r_{\ast\ast}\rho_A}{\rho_M - \rho_B},
\end{sistema}
\;\;\;\;\;\text{that is,}\;\;\;\;\;
\begin{sistema}
	\medskip 
	\rho_M \leq\dfrac{r_{\ast}}{r_{\ast\ast}}\cdot\dfrac{\rho_B\mu_A}{\hat{\mu} - \mu_B} + \rho_A\\
	\medskip
	\rho_M \geq \dfrac{r_{\ast\ast}}{r_{\ast}}\cdot\dfrac{\rho_A\mu_B}{\hat{\mu} - \mu_A} + \rho_B\\
	\rho_M \leq \dfrac{r_{\ast\ast}}{r_{\ast}}\cdot\dfrac{\rho_A(\hat{\mu} - \mu_B)}{\mu_A} + \rho_B,
\end{sistema}
\]
which can be summarized as follows:
\[
\dfrac{r_{\ast\ast}}{r_{\ast}}\cdot\dfrac{\rho_A\mu_B}{\hat{\mu} - \mu_A} + \rho_B\leq \rho_M \leq\min\left\{\dfrac{r_{\ast}}{r_{\ast\ast}}\cdot\dfrac{\rho_B\mu_A}{\hat{\mu} - \mu_B} + \rho_A,\,\dfrac{r_{\ast\ast}}{r_{\ast}}\cdot\dfrac{\rho_A(\hat{\mu} - \mu_B)}{\mu_A} + \rho_B\right\},
\]
or, equivalently,
\[
\max\left\{- \dfrac{r_{\ast}}{r_{\ast\ast}}\cdot\dfrac{\rho_B\mu_A}{\hat{\mu} - \mu_B} - \rho_A,\,- \dfrac{r_{\ast\ast}}{r_{\ast}}\cdot\dfrac{\rho_A(\hat{\mu} - \mu_B)}{\mu_A} - \rho_B\right\}\leq - \rho_M \leq - \dfrac{r_{\ast\ast}}{r_{\ast}}\cdot\dfrac{\rho_A\mu_B}{\hat{\mu} - \mu_A} - \rho_B.
\]
To obtain the condition about the overall risk, we sum $\rho_A + \rho_B$:
\[
\max\left\{\rho_B - \dfrac{r_{\ast}}{r_{\ast\ast}}\cdot\dfrac{\rho_B\mu_A}{\hat{\mu} - \mu_B} ,\,\rho_A - \dfrac{r_{\ast\ast}}{r_{\ast}}\cdot\dfrac{\rho_A(\hat{\mu} - \mu_B)}{\mu_A} \right\}\leq \rho_A + \rho_B - \rho_M \leq \rho_A - \dfrac{r_{\ast\ast}}{r_{\ast}}\cdot\dfrac{\rho_A\mu_B}{\hat{\mu} - \mu_A},
\]
that is,
\[
\begin{aligned}
	\max &\left\{\rho_B\left(1 - \dfrac{r_{\ast}}{r_{\ast\ast}}\cdot\dfrac{\mu_A}{\hat{\mu} - \mu_B}\right),\,\rho_A\left(1 - \dfrac{r_{\ast\ast}}{r_{\ast}}\cdot\dfrac{\hat{\mu} - \mu_B}
	{\mu_A}\right) \right\}\leq \rho_A + \rho_B - \rho_M \\
	&\leq \rho_A\left(1 - \dfrac{r_{\ast\ast}}{r_{\ast}}\cdot\dfrac{\mu_B}{\hat{\mu} - \mu_A}\right).
\end{aligned}
\]
In Case 2, the intersection is the interval $\mathcal{BR}_{\mu}\left(\hat{\mu}\right) = \left[\underline r_{\mu}(\hat{\mu}),\overline r_{\mu}(\hat{\mu})\right]$.    This is represented, under Kulpa's approach, by the point
\[
(x,y)  = \left(\frac{\underline r_{\mu}(\hat{\mu}) + \overline r_{\mu}(\hat{\mu})}{2},\,\frac{\overline r_{\mu}(\hat{\mu}) - \underline r_{\mu}(\hat{\mu})}{2}\right).
\]
This point is unique, that is, the bargaining region does not depend on $\rho_M$. The conditions under which this situation is possible are
\[
\begin{sistema}
	\smallskip\underline r_{\mu}(\hat{\mu})\geq \underline r_{\rho}(\rho_M)\\
	\overline r_{\mu}(\hat{\mu})\leq \overline r_{\rho}(\rho_M),
\end{sistema}
\]
that is,
\[
\begin{sistema}
	\medskip
	\dfrac{r_{\ast}\mu_A}{\hat{\mu} - \mu_B}\geq \dfrac{r_{\ast\ast}(\rho_M - \rho_A)}{\rho_B}\\
	\dfrac{r_{\ast}(\hat{\mu} - \mu_A)}{\mu_B}\leq \dfrac{r_{\ast\ast}\rho_A}{\rho_M - \rho_B},
\end{sistema}
\;\;\;\;\;\text{that is,}\;\;\;\;\;
\begin{sistema}
	\medskip 
	\rho_M \leq\dfrac{r_{\ast}}{r_{\ast\ast}}\cdot\dfrac{\rho_B\mu_A}{\hat{\mu} - \mu_B} + \rho_A\\
	\rho_M \leq \dfrac{r_{\ast\ast}}{r_{\ast}}\cdot\dfrac{\rho_A\mu_B}{\hat{\mu} - \mu_A} + \rho_B,
\end{sistema}
\]
which can be summarized as follows:
\[
\max\{\rho_A,\rho_B\}< \rho_M \leq\min\left\{\dfrac{r_{\ast}}{r_{\ast\ast}}\cdot\dfrac{\rho_B\mu_A}{\hat{\mu} - \mu_B} + \rho_A,\,\dfrac{r_{\ast\ast}}{r_{\ast}}\cdot\dfrac{\rho_A\mu_B}{\hat{\mu} - \mu_A} + \rho_B\right\},
\]
or, equivalently,
\[
\max\left\{- \dfrac{r_{\ast}}{r_{\ast\ast}}\cdot\dfrac{\rho_B\mu_A}{\hat{\mu} - \mu_B} - \rho_A,\,- \dfrac{r_{\ast\ast}}{r_{\ast}}\cdot\dfrac{\rho_A\mu_B}{\hat{\mu} - \mu_A} - \rho_B\right\}\leq - \rho_M <\min\{- \rho_A,-\rho_B\}.
\]
To obtain the condition about the overall risk, we sum $\rho_A + \rho_B$:
\[
\max\left\{\rho_B - \dfrac{r_{\ast}}{r_{\ast\ast}}\cdot\dfrac{\rho_B\mu_A}{\hat{\mu} - \mu_B},\,\rho_A - \dfrac{r_{\ast\ast}}{r_{\ast}}\cdot\dfrac{\rho_A\mu_B}{\hat{\mu} - \mu_A} \right\}\leq\rho_A + \rho_B - \rho_M <\min\{\rho_A,\rho_B\},
\]
that is,
\[
\max\left\{\rho_B\left(1 - \dfrac{r_{\ast}}{r_{\ast\ast}}\cdot\dfrac{\rho_B\mu_A}{\hat{\mu} - \mu_B} \right),\,\rho_A\left(1 - \dfrac{r_{\ast\ast}}{r_{\ast}}\cdot\dfrac{\rho_A\mu_B}{\hat{\mu} - \mu_A} \right)\right\}\leq\rho_A + \rho_B - \rho_M <\min\{\rho_A,\rho_B\}.
\]
In Case 3, the intersection is the interval $\mathcal{BR}_{\rho}\left(\rho_M\right) = \left[\underline r_{\rho}(\rho_M),\overline r_{\rho}(\rho_M)\right]$.    This is represented, under Kulpa's approach, by the point
\[
(x,y)  = \left(\frac{\underline r_{\rho}(\rho_M) + \overline r_{\rho}(\rho_M)}{2},\,\frac{\overline r_{\rho}(\rho_M) - \underline r_{\rho}(\rho_M)}{2}\right),
\]
which, varying $\rho_M$, leads to the curve $\delta$ (see \ref{App2}). The conditions under which this situation is possible are
\[
\begin{sistema}
	\smallskip\underline r_{\mu}(\hat{\mu})\leq \underline r_{\rho}(\rho_M)\\
	\overline r_{\mu}(\hat{\mu})\geq \overline r_{\rho}(\rho_M),
\end{sistema}
\]
that is,
\[
\begin{sistema}
	\medskip
	\dfrac{r_{\ast}\mu_A}{\hat{\mu} - \mu_B}\leq \dfrac{r_{\ast\ast}(\rho_M - \rho_A)}{\rho_B}\\
	\dfrac{r_{\ast}(\hat{\mu} - \mu_A)}{\mu_B}\geq \dfrac{r_{\ast\ast}\rho_A}{\rho_M - \rho_B},
\end{sistema}
\;\;\;\;\;\text{that is,}\;\;\;\;\;
\begin{sistema}
	\medskip 
	\rho_M \geq\dfrac{r_{\ast}}{r_{\ast\ast}}\cdot\dfrac{\rho_B\mu_A}{\hat{\mu} - \mu_B} + \rho_A\\
	\rho_M \geq \dfrac{r_{\ast\ast}}{r_{\ast}}\cdot\dfrac{\rho_A\mu_B}{\hat{\mu} - \mu_A} + \rho_B,
\end{sistema}
\]
which can be summarized as follows:
\[
\rho_M \geq\max\left\{\dfrac{r_{\ast}}{r_{\ast\ast}}\cdot\dfrac{\rho_B\mu_A}{\hat{\mu} - \mu_B} + \rho_A,\,\dfrac{r_{\ast\ast}}{r_{\ast}}\cdot\dfrac{\rho_A\mu_B}{\hat{\mu} - \mu_A} + \rho_B\right\},
\]
or, equivalently,
\[
- \rho_M \leq\min\left\{- \dfrac{r_{\ast}}{r_{\ast\ast}}\cdot\dfrac{\rho_B\mu_A}{\hat{\mu} - \mu_B} -\rho_A,\,- \dfrac{r_{\ast\ast}}{r_{\ast}}\cdot\dfrac{\rho_A\mu_B}{\hat{\mu} - \mu_A} - \rho_B\right\}.
\]
To obtain the condition about the overall risk, we sum $\rho_A + \rho_B$:
\[
\rho_A + \rho_B - \rho_M \leq\min\left\{\rho_B - \dfrac{r_{\ast}}{r_{\ast\ast}}\cdot\dfrac{\rho_B\mu_A}{\hat{\mu} - \mu_B} ,\,\rho_A - \dfrac{r_{\ast\ast}}{r_{\ast}}\cdot\dfrac{\rho_A\mu_B}{\hat{\mu} - \mu_A} \right\},
\]
that is,
\[
\rho_A + \rho_B - \rho_M \leq\min\left\{\rho_B\left(1 - \dfrac{r_{\ast}}{r_{\ast\ast}}\cdot\dfrac{\mu_A}{\hat{\mu} - \mu_B}\right) ,\,\rho_A\left(1 - \dfrac{r_{\ast\ast}}{r_{\ast}}\cdot\dfrac{\mu_B}{\hat{\mu} - \mu_A}\right) \right\}.
\]
In Case 4, the intersection is the interval $\mathcal{CR}_A\left(\hat{\mu},\rho_M\right) = \left[\underline r_{\rho}(\rho_M),\overline r_{\mu}(\hat{\mu})\right]$. This is represented, under Kulpa's approach, by the point
\[
(x,y)  = \left(\frac{\underline r_{\rho}(\rho_M) + \overline r_{\mu}(\hat{\mu})}{2},\,\frac{\overline r_{\mu}(\hat{\mu}) - \underline r_{\rho}(\rho_M)}{2}\right),
\]
that is,
\[
x = \frac{1}{2}\left(\frac{r_{\ast\ast}(\rho_M - \rho_A)}{\rho_B} + \frac{r_{\ast}(\hat{\mu} - \mu_A)}{\mu_B}\right),
\]
\[
y = \frac{1}{2}\left(\frac{r_{\ast}(\hat{\mu} - \mu_A)}{\mu_B} - \frac{r_{\ast\ast}(\rho_M - \rho_A)}{\rho_B}\right).
\]
The sum of its coordinates is
\[
x + y = \frac{r_{\ast}(\hat{\mu} - \mu_A)}{\mu_B};
\]
that is, on varying $\rho_M$ we move along the straight line of equation
\[
y = - x + \frac{r_{\ast}(\hat{\mu} - \mu_A)}{\mu_B}.
\]
The conditions under which this situation is possible are
\[
\begin{sistema}
	\smallskip\underline r_{\mu}(\hat{\mu})\leq \underline r_{\rho}(\rho_M)\\
	\smallskip\overline r_{\mu}(\hat{\mu})\leq \overline r_{\rho}(\rho_M)\\
	\underline r_{\rho}(\rho_M)\leq \overline r_{\mu}(\hat{\mu}),
\end{sistema}
\]
that is,
\[
\begin{sistema}
	\medskip
	\dfrac{r_{\ast}\mu_A}{\hat{\mu} - \mu_B}\leq \dfrac{r_{\ast\ast}(\rho_M - \rho_A)}{\rho_B}\\
	\medskip
	\dfrac{r_{\ast}(\hat{\mu} - \mu_A)}{\mu_B}\leq \dfrac{r_{\ast\ast}\rho_A}{\rho_M - \rho_B}\\
	\dfrac{r_{\ast\ast}(\rho_M - \rho_A)}{\rho_B}\leq\dfrac{r_{\ast}(\hat{\mu} - \mu_A)}{\mu_B},
\end{sistema}
\;\;\;\;\;\text{that is,}\;\;\;\;\;
\begin{sistema}
	\medskip 
	\rho_M \geq\dfrac{r_{\ast}}{r_{\ast\ast}}\cdot\dfrac{\rho_B\mu_A}{\hat{\mu} - \mu_B} + \rho_A\\
	\medskip
	\rho_M \leq \dfrac{r_{\ast\ast}}{r_{\ast}}\cdot\dfrac{\mu_B\rho_A}{\hat{\mu} - \mu_A} + \rho_B\\
	\rho_M \leq \dfrac{r_{\ast}}{r_{\ast\ast}}\cdot\dfrac{\rho_B(\hat{\mu} - \mu_A)}{\mu_B} + \rho_A,
\end{sistema}
\]
which can be summarized as follows:
\[
\dfrac{r_{\ast}}{r_{\ast\ast}}\cdot\dfrac{\rho_B\mu_A}{\hat{\mu} - \mu_B} + \rho_A\leq \rho_M \leq\min\left\{\dfrac{r_{\ast\ast}}{r_{\ast}}\cdot\dfrac{\mu_B\rho_A}{\hat{\mu} - \mu_A} + \rho_B,\,\dfrac{r_{\ast}}{r_{\ast\ast}}\cdot\dfrac{\rho_B(\hat{\mu} - \mu_A)}{\mu_B} + \rho_A\right\},
\]
or, equivalently,
\[
\max\left\{- \dfrac{r_{\ast\ast}}{r_{\ast}}\cdot\dfrac{\mu_B\rho_A}{\hat{\mu} - \mu_A} - \rho_B,\,- \dfrac{r_{\ast}}{r_{\ast\ast}}\cdot\dfrac{\rho_B(\hat{\mu} - \mu_A)}{\mu_B} - \rho_A\right\}\leq - \rho_M \leq - \dfrac{r_{\ast}}{r_{\ast\ast}}\cdot\dfrac{\rho_B\mu_A}{\hat{\mu} - \mu_B} - \rho_A.
\]
To obtain the condition about the overall risk, we sum $\rho_A + \rho_B$:
\[
\max\left\{\rho_A - \dfrac{r_{\ast\ast}}{r_{\ast}}\cdot\dfrac{\mu_B\rho_A}{\hat{\mu} - \mu_A},\,\rho_B - \dfrac{r_{\ast}}{r_{\ast\ast}}\cdot\dfrac{\rho_B(\hat{\mu} - \mu_A)}{\mu_B} \right\}\leq \rho_A + \rho_B - \rho_M \leq \rho_B - \dfrac{r_{\ast}}{r_{\ast\ast}}\cdot\dfrac{\rho_B\mu_A}{\hat{\mu} - \mu_B} ,
\]
that is,
\[
\begin{aligned}
	\max &\left\{\rho_A\left(1 - \dfrac{r_{\ast\ast}}{r_{\ast}}\cdot\dfrac{\mu_B}{\hat{\mu} - \mu_A}\right),\,\rho_B\left(1 - \dfrac{r_{\ast}}{r_{\ast\ast}}\cdot\dfrac{\hat{\mu} - \mu_A}
	{\mu_B}\right) \right\}\leq \rho_A + \rho_B - \rho_M \\
	&\leq \rho_B\left(1 - \dfrac{r_{\ast}}{r_{\ast\ast}}\cdot\dfrac{\mu_A}{\hat{\mu} - \mu_B}\right).
\end{aligned}
\]

\bibliography{FusioniBib}{}
\bibliographystyle{ieeetr}

\end{document}